\documentclass[a4paper,11pt]{article}

\usepackage{jcappub}
\usepackage{journal_abbrev}
\usepackage{xstring}
\usepackage{xparse}
\usepackage{xspace}
\usepackage{etoolbox}
\xspaceaddexceptions{]\}}
  \usepackage{bm}

  \usepackage{hyperref}

\newcommand{\twofig}[4]{%
  \begin{figure*}%
    \centerline{\resizebox{\hsize}{!}{\includegraphics*{#1} \,%
        \includegraphics*{#2}}}%
    \caption{#4}\label{#3}%
  \end{figure*}%
}

\DeclareDocumentCommand{\eqs}{m m m o o}{%
  \IfNoValueTF {#4} {%
    Eqs.~(\ref{#1}){\xspace #2} (\ref{#3})%
  }{%
    Eqs.~(\ref{#1}){\xspace #2} (\ref{#3}){\xspace #4} (\ref{#5})%
  }%
}

\newenvironment{referee}{\bf}{}
\newcommand{\bref}{\begin{referee}}
\newcommand{\eref}{\end{referee}}

\def\ba#1\ea{\begin{align}#1\end{align}}
\def\bea{\begin{eqnarray}}
\def\eea{\end{eqnarray}}
\def\be{\begin{equation}}
\def\ee{\end{equation}}
\def\d{\delta}
\def\s{\sigma}
\def\({\left(}
\def\){\right)}
\def\[{\left[}
\def\]{\right]}

\def\<{\left\langle}
\def\>{\right\rangle}
\def\lapl{\nabla^2}
\DeclareMathOperator{\tr}{tr}
\DeclareMathOperator{\diag}{diag}
\def\shouldeq{=}

\newcommand{\vs}{\nonumber\\}
\newcommand{\cs}{\,,\  } 
\def\d{{\delta}}
\def\eps{{\varepsilon}}
\def\LO{\text{LO}}
\def\NLO{\text{NLO}}
\def\otd{\text{td}}

\newcommand{\perm}[1]{ \expandafter\ifstrempty\expandafter{#1} {\mbox{perm.}} {\mbox{$#1$ perm.}} }

\def\thH{\Theta_\text{H}}

\renewcommand{\v}[1]{\bm{#1}}

\def\vx{\v{x}}

\def\vk{\v{k}}
\def\vq{\v{q}}
\def\vp{\v{p}}

\def\O{\mathcal{O}}
\def\Oset{\mathfrak{O}}
\def\Del{\mathcal{D}}
\def\rhob{\overline{\rho}_m}
\def\avng{\overline{n}_h}
\def\vkhat{\hat{\vk}}
\def\knl{k_\text{NL}}
\def\xfl{\vx_\text{fl}}
\def\clapl{c_{\lapl\d}}

\newcommand{\norm}[1]{\left\lVert#1\right\rVert}

\def\Mpch{\,h^{-1}\text{Mpc}}
\def\iMpch{\,h\,\text{Mpc}^{-1}}
\def\Plin{P_\text{L}}
\def\Om{\Omega_m}
\def\cH{\mathcal{H}}
\def\Lbox{L_\text{box}}
\def\L{\Lambda} 

\def\fsum#1{\sum_{#1\neq 0}^{k_{\rm max}}}

\newcommand{\refeq}[1]{Eq.~(\ref{eq:#1})}

\newcommand{\reffig}[1]{Fig.~\ref{fig:#1}}

\newcommand{\reftab}[1]{Tab.~\ref{tab:#1}}
\newcommand{\refsec}[1]{Sec.~\ref{sec:#1}}

\newcommand{\refapp}[1]{Appendix~\ref{app:#1}}

\title{A rigorous EFT-based forward model for large-scale structure}

\author[a]{Fabian Schmidt,}
\author[a]{Franz Elsner,}
\author[b]{Jens Jasche,}
\author[a]{Nhat~Minh~Nguyen,}
\author[c]{Guilhem Lavaux}

\emailAdd{fabians@mpa-garching.mpg.de}
\emailAdd{felsner@mpa-garching.mpg.de}
\emailAdd{jens.jasche@fysik.su.se}
\emailAdd{minh@mpa-garching.mpg.de}
\emailAdd{guilhem.lavaux@iap.fr}

\affiliation[a]{Max--Planck--Institut f\"ur Astrophysik,
  Karl--Schwarzschild--Stra\ss e 1, D--85748 Garching, Germany}
\affiliation[b]{The Oskar Klein Centre, Department of Physics, Stockholm University, Albanova University Center, SE 106 91 Stockholm, Sweden}
\affiliation[c]{Sorbonne Universit\'e, CNRS, UMR 7095, Institut d'Astrophysique de Paris, 98 bis bd Arago, 75014 Paris, France}


\abstract{
  Conventional approaches to cosmology inference from galaxy redshift surveys are based on $n$-point functions, which are under rigorous perturbative control on sufficiently large scales. Here, we present an alternative approach, which employs a likelihood at the level of the galaxy density field. By integrating out small-scale modes based on effective-field theory arguments, we prove that this likelihood is under perturbative control if certain specific conditions are met. We further show that the information captured by this likelihood is equivalent to the combination of the next-to-leading order galaxy power spectrum, leading-order bispectrum, and BAO reconstruction. Combined with MCMC sampling and MAP optimization techniques, our results allow for fully Bayesian cosmology inference from large-scale structure that is under perturbative control. We illustrate this via a first demonstration of unbiased cosmology inference from nonlinear large-scale structure using this likelihood. In particular, we show unbiased estimates of the power spectrum normalization $\sigma_8$ from a catalog of simulated dark matter halos, where nonlinear information is crucial in breaking the $b_1-\sigma_8$ degeneracy.
  }

\begin{document}

\maketitle

\flushbottom

\section{Introduction}
\label{sec:intro}
One of the prime difficulties in inferring cosmology from the observed large-scale structure is the nonlinear and nonlocal connection between the matter density and tidal field, whose evolution is well understood, and the density of observed tracers such as galaxies. One possible approach, which offers the advantages of theoretical robustness as well as controlled systematic uncertainties is the \emph{effective field theory (EFT) approach}, or equivalently, \emph{general bias expansion} (see \cite{baumann/etal:2012,carrasco/etal:2012} for details on the EFT approach in large scale structure, and \cite{biasreview} for a review of bias). In this approach, the fully nonlinear small-scale modes are integrated out, leading to a relation between the observed galaxy density and operators constructed out of the matter density field multiplied by free coefficients, the bias parameters. Predictions are obtained by truncating the expansion at fixed order in perturbations and spatial derivatives. This approach is inherently limited to scales larger than the scale where the matter density field becomes nonlinear, at a wavenumber $\knl \sim 0.25 \iMpch$ for a standard $\Lambda$CDM cosmology at today's epoch, and larger than the characteristic spatial scale $R_*$ involved in the formation of the tracer considered.

Standard techniques for applying this approach to observational or
simulated data sets make use of summary statistics, in particular $n$-point correlation functions in real or Fourier space on large scales (large $r$, small $k$), where $n \geq 2$  (see Sec.~4.1 of \cite{biasreview} for an overview). However, in order to extract information from galaxy statistics beyond the linear regime in this approach, higher-order $n$-point functions, such as the three- and four-point functions, are crucial. By combining 2- and 3-point functions, for example, the degeneracy between linear bias and the amplitude of matter fluctuations can be broken, allowing for cosmological constraints.

Unfortunately, both estimators and the associated covariances for higher-order $n$-point functions become increasingly difficult to handle with growing $n$,
  both technically and computationally, due to the rapidly increasing size of the data vectors. In particular, survey-specific systematic effects such as the mask, varying survey depth, or fiber-collision effects, need to be incorporated into the model for each $n$-point function. A possible alternative is to attempt to forward-model the galaxy density field itself, without resorting to summary statistics. This approach offers the advantage of a much more straightforward incorporation of systematic effects.
  Starting from early attempts based on galaxy peculiar velocities
\cite{
  1999ApJ...522....1D,
  1992ApJ...391..443N},
this approach is being pursued increasingly actively 
\cite{
  2008MNRAS.389..497K,
  2010MNRAS.407...29J,
  2010MNRAS.409..355J,
  2013MNRAS.432..894J,
  2013ApJ...772...63W,
  2013ApJ...779...15J,
  2014ApJ...794...94W,
  2015MNRAS.446.4250A,
  2017JCAP...12..009S,
  modi/etal,
  2018arXiv180611117J}. The forward model for matter, together with the perturbative bias expansion, provides us with a ``mean tracer field'' in a certain sense. But which likelihood should be used to compare this mean field with the observed galaxy density field? 
Answering this question, on which little theoretical progress has been made so far, in the context of the EFT is the goal of this paper. The studies cited above assumed localized likelihoods in real space either of a specific form, or modeled using neural networks. As we will see, the EFT approach arrives at a somewhat different result.

After reviewing the program of Bayesian LSS inference, we derive the likelihood in the EFT approach by integrating out small-scale modes. As we will see, this yields an approximately Gaussian likelihood in Fourier space. We restrict to the rest-frame density of a tracer, i.e. neglect redshift-space distortions. Hence, even though our results in principle apply to any LSS tracer, we will refer to the tracer as ``halos'' throughout the paper, as the most straightforward application consists of halos identified in N-body simulations. We further derive and investigate the maximum-likelihood point when phases are fixed, and show that it corresponds to matching the halo-matter cross-power spectrum at next-to-leading (1-loop) order, and the halo-matter-matter bispectrum at leading order (tree level). We derive the precise conditions that need to be met to ensure an unbiased result.

Finally, we show that this approach naturally includes fully nonlinear reconstruction of the baryon acoustic oscillation (BAO) feature.
BAO reconstruction refers to the fact that the displacements of galaxies from their initial (Lagrangian) positions lead to a damping of the oscillatory BAO feature. These displacements are dominated by large-scale modes which are inferred jointly with the cosmological parameters when following a forward-modeling approach. This offers another clear advantage over approaches based on $n$-point functions.

The outline of the paper is as follows. \refsec{forward} reviews the Bayesian posterior for large-scale structure, focusing on large scales, which involves the prior on the initial conditions, forward model for matter, and deterministic bias expansion. \refsec{GaussF} then derives the remaining missing ingredient, the conditional probability of the observed halo field given the final matter density field and bias parameters. The following sections examine the ramifications of this likelihood: in \refsec{PT}, we study the maximum-likelihood point of this likelihood, and derive its relation to $n$-point correlation functions in the EFT. \refsec{nongauss} shows how non-Gaussian corrections to the likelihood are suppressed. Finally, \refsec{reco} describes how the likelihood presented here incorporates BAO reconstruction. We then turn to a preview of numerical results based on this likelihood in \refsec{results}, which will be described in an upcoming publication \cite{paperII}, and conclude in \refsec{concl}.
The appendices contain colletions of important relations as well as details on the calculations presented in the main text.

\subsection*{Notation}

Our notation largely follows that of \cite{biasreview}. In particular, our Fourier convention and short-hand notation is
\ba
f(\vk) \equiv\:& \int d^3 \vx\, f(\vx) e^{-i\vk\cdot\vx} \equiv \int_{\vx} f(\vx) e^{-i\vk\cdot\vx} \vs
f(\vx) \equiv\:& \int \frac{d^3 \vk}{(2\pi)^3}\, f(\vk) e^{i\vk\cdot\vx} \equiv \int_{\vk} f(\vk) e^{i\vk\cdot\vx}\,.
\ea
Wavenumbers that are integrated over (loop momenta) will further be denoted as $\vp, \vp',\vp_i\cdots$. 
Primes on Fourier-space correlators indicate that the momentum conserving Dirac delta $(2\pi)^3 \d_D(\vk_1+\vk_2+\cdots)$ is to be dropped.
We will also frequently use the nonlocal derivative operator
\be
\Del_{ij} \equiv \left(\partial_i\partial_j \nabla^{-2} - \frac13 \d_{ij} \right)
\label{eq:Deldef}
\ee
which is defined via its action on fields in the Fourier representation.
We will typically deal with the filtered density field, employing an unspecified smoothing kernel $W(\vx)$ with Fourier-space counterpart $W(\vk)$. We will further use $W_{\L}$ for a sharp-$k$ filter:
\be
W_{\L}(\vk) = \thH(\Lambda-|\vk|)\,.
\ee
The matter and rest-frame halo (or galaxy) density perturbations are given by
\be
\d(\vx,\tau) \equiv \frac{\rho(\vx,\tau) - \rhob(\tau)}{\rhob(\tau)} \quad\mbox{and}\quad
\d_h(\vx,\tau) \equiv \frac{n_h(\vx,\tau) - \avng(\tau)}{\avng(\tau)}\,.
\ee
Correspondingly, we denote the filtered matter and halo fields as $\d_W$ and $\d_{h,W}$, respectively. We reserve the notation $\d_\L,\  \d_{h,\L}$ for fields filtered with a sharp-$k$ filter on the scale $\L$.

Since the matter density is related to the potential $\Phi$ through the Poisson equation
\be
\lapl\Phi = \frac32 \Om \cH^2 \d\,,
\ee
this allows us to combine the matter density perturbation and tidal field $K_{ij}$ into a tensor $\Pi^{[1]}$:
\be
\Pi^{[1]}_{ij}(\vx,\tau) \equiv \frac{2}{3\Om\cH^2} \partial_{x,i}\partial_{x,j}\Phi(\vx,\tau)
=
K_{ij}(\vx,\tau) + \frac13 \delta_{ij}\d(\vx,\tau)\,,
\label{eq:Pi1}
\ee
which contains $\d = \tr \Pi^{[1]}$ and $K_{ij}$ as the trace-free part of
$\Pi^{[1]}_{ij}$. All of these quantities are the evolved, nonlinear quantities.
We further use the following notation for perturbative order:
\begin{itemize}
\item[$O^{(n)}\,:$] Operator evaluated at $n$-th order in perturbation theory.
\item[$O^{[n]}\,:$] Operator whose \emph{lowest-order} contribution is at $n$-th order in perturbation theory.
\end{itemize}
All fields are implicitly assumed to be filtered on the grid scale used in the forward model. We will denote the filtering explicitly only when it is contrasted with unfiltered fields (specifically in \refsec{PT}).

Finally, we let $\vec{\d}$ stand for a field defined on a grid: $\vec{\d}=\{ \d(\vx_i) \}_{i=1}^{N_g^3}$, where $N_g$ is the number of grid cells on one side, and throughout latin indices $i,j,k,...$ label grid cells, while greek indices $\alpha, \beta,...$ label parameters. The set of cosmological parameters will usually be denoted as $\theta$, while we reserve $\lambda$ for ``nuisance'' parameters (e.g. moments of the likelihood).
We will also frequently use the corresponding discrete fields defined in the Fourier domain. For this, we adopt the standard box normalization:
\ba
\d(\vk) &= \sum_i^{N_g^3} \d(\vx_i) e^{-i \vk\cdot \vx_i}
\label{eq:boxnorm}\\
\d(\vx) &= \frac1{N_g^3}\sum_{\vk}^{k_{\rm Ny}} \d(\vk_i) e^{i\vk_i\cdot \vx}
\quad\mbox{where}\quad
\vk \in (n_x, n_y, n_z) k_F\,,\  k_F = \frac{2\pi}{\Lbox}\,,
\nonumber
\ea
and $n_i \in \{-N_g/2, \cdots N_g/2\}$ while $k_{\rm Ny} \equiv N_g k_F/2$. We will often use
\be
\fsum{\vk} \equiv \sum_{\{n_x,n_y,n_z\} \neq \{0,0,0\}}^{n_x^2+n_y^2+n_z^2 \leq  (k_{\rm max}/k_F)^2}\,.
\ee
Correlators of box-normalized fields obey
\ba
\< X(\v{n} k_F) Y(\v{n}' k_F) \> &= \frac1{\Lbox^3} \d_{\v{n},-\v{n}'} P_{XY}(\v{n} k_F) \vs
\< X(\v{n} k_F) Y(\v{n}' k_F) Z(\v{n}'' k_F) \> &= \frac1{\Lbox^6} \d_{\v{n}+\v{n}',-\v{n}''}  B_{XYZ}(\v{n} k_F, \v{n}' k_F) \,,
\label{eq:corrbox}
\ea
where
\be
\d_{\v{n},\v{n}'} \equiv \d_{n_x n_x'}\d_{n_y n_y'}\d_{n_z n_z'}\,,
\ee
and $P_{XY}$ ($B_{XYZ}$) are the cross-power spectrum (bispectrum) respectively. We have neglected the averaging over the finite $k$-space bin of width $k_{\rm Ny}$ on the r.h.s. of \refeq{corrbox}.

\section{Bayesian forward model}
\label{sec:forward}

Our goal is to derive a joint posterior for the initial density field $\vec{\d}_\text{in}$, cosmological parameters $\theta$, and ``nuisance parameters,'' i.e. bias parameters and stochastic amplitudes, which describe the uncertainties in the halo formation process. This posterior involves four ingredients:
\begin{enumerate}
\item The prior on the initial conditions.
\item The forward model for matter and gravity.
\item The deterministic bias model.
\item The conditional likelihood for the halo density at a given point.
\end{enumerate}
The first ingredient essentially corresponds to specifying the cosmological model space.
As we will see, given the well-developed understanding of the second \cite{bernardeau/etal:2002} and third ingredients \cite{biasreview}, the last point is the core open issue in this enterprise.

Throughout, we assume for simplicity that we are considering a single tracer field $\vec{\d}_h$ at a fixed time $\tau$, which we will leave implicit. The former assumptions can be generalized straightforwardly.

\subsection{Prior on initial conditions}
\label{sec:forward:prior}

We assume adiabatic, growing-mode initial conditions, so that the initial conditions are given by a single field $\vec{\d}_\text{in}$, the initial density field. We further assume a multivariate Gaussian distribution as prior on $\vec{\d}_\text{in}$, determined by the linear matter power spectrum $\Plin(k,\theta)$, which depends on the set of cosmological parameters $\theta$:
\be
P_\text{prior}\left(\vec{\d}_\text{in}|\theta\right) = \mathcal{N}\left(\vec{\d}_\text{in}\Big|\ \vec{\mu}= \vec{0},\  \text{C}=\text{FT}^\dagger [\diag\{\Plin(k_i,\theta)\}] \text{FT} \right)\,,
\label{eq:Pprior}
\ee
where $\mathcal{N}$ denotes a multivariate Gaussian distribution, $\vec{\mu}$ is the vanishing expectation value and $\text{C}$ is the covariance matrix, which is diagonal in Fourier space.
This can of course be generalized to include primordial non-Gaussianity, and isocurvature perturbations between baryons and CDM. We will briefly comment on these in \refsec{concl}.

\subsection{Forward model for matter and gravity}

The forward model of gravity yields the probability of finding an evolved density field $\vec{\d}$, given an initial density field $\vec{\d}_\text{in}$ and cosmological parameters $\theta$. We will write this as a deterministic model (see also \cite{2018arXiv180611117J}):
\be
P\left(\vec{\d}\,\Big|\vec{\d}_\text{in},\theta\right) = \prod_{i=1}^{N_g^3} \d_D\left(\d^i - \d_\text{fwd}^i[\vec{\d}_\text{in}, \theta]\right)\,.
\label{eq:Pfwd}
\ee
The actual physics of the forward model, e.g. Lagrangian perturbation theory or particle-mesh simulation, is encoded in the nonlinear nonlocal functional $\d_\text{fwd}^i[\vec{\d}_\text{in}, \theta]$.
Any such forward model $\d_\text{fwd}^i[\vec{\d}_\text{in}, \theta]$ will be imperfect of course, both due to approximations made in the calculation and the fact that only modes down to some finite minimum scale are included, a fact which is neglected in \refeq{Pfwd}. We will return to this below in the context of the conditional likelihood.

Specifically, as long as certain conditions regarding mass and momentum conservation of matter are met, the small-scale uncertainties in the forward model for matter and gravity can be effectively included in the conditional likelihood. In the EFT context, the main requirement is that the forward model consistently includes all relevant terms up to a fixed order in perturbation theory, and that numerical truncation errors can be neglected. To be specific, in the main text we will assume that the error in the forward model of matter is at least third order in perturbation theory. The results from our practical implementation shown in \refsec{results} are based on 2LPT. Nothing however---apart from computational expense---prevents the approach presented here from being coupled to a full numerical simulation as forward model.

\subsection{Bias expansion}
\label{sec:bias}

We now define the ``deterministic'' halo density field at a given point as a linear superposition of operators, or fields, $O$,
\be
\vec{\d}_{h,\rm det}[\vec{\d},\{b_O\}] = \sum_O b_O \vec{O}[\vec{\d}]\,,
\label{eq:dhdet}
\ee
where again $\vec{\d}$ denotes the evolved matter density field, and the $\vec{O}$ are constructed to encompass the complete linearly independent set of local gravitational observables at a certain order in perturbation theory: matter density and velocity divergence, tidal field, and so on. At a given order in perturbation theory, there is only a finite  fixed number of linearly independent gravitational observables \cite{senatore:2014,MSZ}. Moreover, even though the gravitational observables include time derivatives, all of these operators can be expressed as nonlocal, nonlinear transformations of the final density field $\vec{\d}$ \cite{MSZ} (see Sec.~2.5 of \cite{biasreview} for a review). $b_O$ are the corresponding bias parameters.

\refeq{dhdet} only predicts the halo density field in a statistical sense. That is, if we imagine stacking many cells that have the same values of all operators $O$ appearing in \refeq{dhdet}, then the mean density of halos in these cells should approach the prediction in \refeq{dhdet}.
At any given point, the halo density can deviate from the prediction in \refeq{dhdet}, due to the random nature of the small-scale perturbations that we have integrated out in the bias expansion, but that yet are relevant for halo formation. This ``scatter'' will be taken into account in the conditional probability which we discuss below.
Further, we reiterate that the operators in \refeq{dhdet} are constructed from the evolved density field on the grid, i.e. \emph{filtered on the grid scale.} This will become relevant later.

In this paper, we use the following default set of operators:
\ba
O &\in \Big\{ \d,\  \d^2 - \<\d^2\>,\  (K_{ij}^2) - \<(K_{ij})^2\>,\  \lapl\d \Big\}\,,
\vs
\mbox{with coefficients}& \quad
\left\{ b_1,\  \frac{b_2}2,\  b_{K^2},\  \clapl \right\}\,.
\label{eq:Osetfid}
\ea
We further define $b_N \equiv N! b_{\d^N}$ as the bias coefficient corresponding to the $N$-th power of the matter density field (local-in-matter-density, LIMD). We denote the higher-derivative bias coefficient as $\clapl$ rather than $b_{\lapl\d}$, as it is an effective coefficient which also absorbs other contributions which depend on the chosen smoothing scale and cutoff, as we will see. More generally, the bias parameters $b_O$ correspond to well-defined physical bias parameters, while parameters denoted as $c_O$ (so far, only $\clapl$) are understood as effective coefficients which also absorb higher-order contributions.

\refeq{Osetfid} corresponds to the complete set of operators up to second order at leading order in derivatives, and the leading higher-derivative operator ($\lapl\d$). The significance of this choice will become clear later.

\subsection{Stochasticity and conditional probability}

The final ingredient needed in the Bayesian forward model is the probability for finding a certain number of halos or galaxies in a given cell, given the predicted deterministic field $\vec{\d}_{h,\rm det}$ (as well as the matter density field). We will phrase this equivalently as the \emph{conditional probability for finding a measured halo density field $\vec{\d}_h$ given the predicted mean-field halo density $\vec{\d}_{h,\rm det}$.} As discussed in the previous section, this probability should take into account the scatter induced by the small-scale modes that are not explicitly included in the forward model, which are nevertheless relevant for determining exactly where a halo forms. Further, the conditional probability also needs to be able to capture deficiencies in the bias expansion \refeq{dhdet}, as well as in the forward model for matter and gravity.

One approach, followed by essentially all literature on this topic so far, is to assume that the size of the grid cells $R_\text{cell}$ is much larger than the scale $R_*$ that controls the higher-derivative contributions to the halo density in \refeq{dhdet}. That is, one assumes that on the scales resolved on the grid, halo formation can be effectively approximated as spatially local. In the EFT approach, the leading correction to this assumption is captured by the operator $\lapl\d$ in \refeq{Osetfid}, whose coefficient then should be (at least) of order $R_*^2$, i.e. $|\clapl| \sim R_*^2$. Alternatively, one can explicitly include the difference between neighboring cells, as done in \cite{modi/etal}. This approach thus assumes that the impact of all modes resolved on the grid, including the correlations of the halo density between different cells, is completely captured by a finite set of operators, in our case those appearing in \refeq{dhdet}.

Thus, building on the assumption on the locality of halo formation, the likelihood of a given halo density field \emph{given} a matter density field and bias parameters is a product of conditional probabilities in each cell:
\be
P\left(\vec{\d}_h\Big|\vec{\d}, \{b_O\}, \{\lambda_a\}\right) = \prod_{i=1}^{N_g^3} P^{(1)}\left(\d_h^i - \d_{h,\rm det}^i[\vec{\d},\{b_O\}],\  \{\lambda_a\},\  \d^i\right)\,,
\label{eq:Pcond}
\ee
where $P^{(1)}$ is the probability for finding, in a given cell, an overdensity $\d_h^i$ given the prediction for the mean relation $\d_{h,\rm det}^i$ from \refeq{dhdet}.  Here we have allowed $P^{(1)}$ to depend on further parameters $\{\lambda_a\}$ (e.g., variance, skewness, ...), as well as the matter density itself to take into account, for example, a larger variance in high-density regions.

Clearly, \refeq{Pcond} still contains significant freedom to choose the form of the conditional probability $P^{(1)}$. Moreover, unlike the case for the bias expansion in \refeq{dhdet}, there is no guide from effective field theory considerations on what the form of $P^{(1)}$ should be, since $P^{(1)}$ arises from integrating out small-scale, fully nonlinear modes whose PDF is not expected to be close to Gaussian. There is a limit for which the PDF is expected to asymptote to a known form: if the size of grid cells is much less than the mean separation between halos, then we expect the single-cell PDF to approach a Poisson distribution by virtue of the law of rare events. Unfortunately, this limit is not attainable within an EFT context however, since the perturbative bias expansion \refeq{dhdet} breaks down for such a small grid scale (in practice, the grid scale will then also be much smaller than $R_*$). Thus, in order to obtain a posterior that is under rigorous perturbative control, we need to pursue a different route than \refeq{Pcond}.

\subsection{Final posterior}

Before proceeding further, let us put together the ingredients presented above, to obtain the final joint posterior of the initial density field, cosmological parameters, and nuisance parameters:
\ba
P\left(\vec{\d}_\text{in},\theta,\{b_O\},\{\lambda_a\}\Big| \vec{\d}_h\right) =\:&
\mathcal{N}_P P_\text{prior}(\vec{\d}_\text{in}|\theta) 
P\left(\vec{\d}_h \Big| \vec{\d}_\text{fwd}[\vec{\d}_\text{in},\theta],\  \{b_O\},\  \{\lambda_a\} \right)\,,
 \label{eq:post}
\ea
where $\mathcal{N}_P$ is a normalization constant. Then, the desired cosmological constraints can be obtained by marginalizing over the initial phases $\vec{\d}_\text{in}$, as well as nuisance parameters:
\ba
P\left(\theta\Big| \vec{\d}_h\right) =\:& \int d\{b_O\} \int d\{\lambda_a\}\
P_\text{prior}(\{b_O\}, \{\lambda_a\}) \int \mathcal{D}\vec{\d}_\text{in}
P\left(\vec{\d}_\text{in},\theta,\{b_O\},\{\lambda_a\}\Big| \vec{\d}_h\right)\,,
\label{eq:postmarg}
\ea
where $P_\text{prior}$ is a prior on the bias and PDF parameters.
Of course, one can similarly obtain marginalized posteriors for the bias parameters $b_O$ and likelihood parameters $\lambda_a$.

\section{EFT approach to the conditional probability}
\label{sec:GaussF}

The conditional probability \refeq{Pcond} is written in real space, i.e. it relates the fields $\vec{\d}_h$ and $\vec{\d}_{h,\rm det}$ cell-by-cell in real space. Perturbative approaches, including the effective field theory of biased tracers, however naturally work in Fourier space. This is because the initial conditions are (approximately) a homogeneous Gaussian random field (see \refsec{forward:prior}), so that their covariance is diagonal in Fourier space.

Let us thus instead consider the problem of deriving a conditional probability relating the halo density field in Fourier space, $\vec{\d}_h \to \d_h(\vk)$, to  a linear combination $\d_{h,\rm det}(\vk)$ of fields $O(\vk)$ constructed from the evolved matter field $\d(\vk)$. Clearly, the field $\d_{h,\rm det}(\vk)$ cannot be a perfect match to $\d_h(\vk)$, but has noise. In the EFT approach, this noise formally arises because we are integrating out the small-scale modes of the density field, those above some maximum wavenumber $k_{\rm max}$.
An obvious question then is how the maximum wavenumber $k_{\rm max}$ should be chosen. A priori, the only guidance we have is that it should be less than the nonlinear scale, $k_{\rm NL} \approx 0.3\iMpch$ at redshift zero (but higher at higher redshifts). We will make this more precise in \refsec{PT:HO} below.

Importantly, the noise does not only affect $\d_h$, but in general the fields entering our predicted field $\d_{h,\rm det}$ also have noise. We can thus write
\be
\d_h(\vk) - \d_{h,\rm det}(\vk) = \eps_h(\vk) - \eps_\text{model}(\vk) = \eps_h(\vk) - b_1 \eps_m(\vk)\,.
\label{eq:epsdef}
\ee
In the second equality, we have only kept the noise contribution from the matter density $\eps_m$, which is multiplied by $b_1$ since this is how the matter density field enters $\d_{h,\rm det}$. The contributions from noise fields in the quadratic operators (specifically those that cannot be absorbed by $\eps_h$) are higher order, as we will show in \refsec{PT:quadrnoise}, and can thus be dropped. Apart from this ranking, we have not used perturbative arguments so far. We will work in the continuum limit to keep the conceptual derivation clear, and move to a finite Fourier grid shortly.

Our goal is now to integrate out the noise fields $\eps_h$ and $\eps_m$, since, by definition, we cannot predict them at the field level. For this, we assume that, on the scales $k < k_{\rm max}$ of interest, both fields can be approximated as Gaussian. On sufficiently large scales, this approximation is guaranteed to be accurate by the central limit theorem. We will return to this point in \refsec{nongauss}. Further, we can use the fact that the power spectra of the noise fields have to be analytic in $k$ on large scales. This is because they arise from interactions of modes that are of much smaller scale, and thus cannot involve the power spectrum on the scale $k$ \cite{abolhasani/mirbabayi/pajer:2016}. Finally, the noise in the matter density field has to satisfy $\lim_{k\to 0} \eps_m(\vk)/k^2 =$~const., through mass and momentum conservation. We thus write, up to including $k^4$,
\ba
\< \eps_h(\vk) \eps_h(\vk)\>' &= P^\eps_{hh}(k) = P^{\eps,0}_{hh} + P^{\eps,2}_{hh} k^2 + P^{\eps,4}_{hh} k^4 \vs
\< \eps_h(\vk) \eps_m(\vk)\>' &= P^\eps_{hm}(k) = P^{\eps,2}_{hm} k^2 + P^{\eps,4}_{hm} k^4 \vs
\< \eps_m(\vk) \eps_m(\vk)\>' &= P^\eps_{mm}(k) = P^{\eps,4}_{mm} k^4 \,.
\ea
The Cauchy-Schwartz inequality implies $|P^{\eps,2}_{hm}|\leq \sqrt{P^{\eps,0}_{hh}P^{\eps,4}_{mm}}$. To summarize, the noise fields $\eps_h,\eps_m$ follow a multivariate normal distribution given by
\ba
P\Big(\eps_h(\vk),\eps_m(\vk)\Big) &= | 2\pi \text{C}_\eps |^{-1/2}
\exp\left[-\frac12 (\eps_h,\  \eps_m ) \text{C}_\eps^{-1} (\eps_h,\  \eps_m )^\dagger \right]\,, \vs
\mbox{where}\quad \text{C}_\eps = \text{C}_\eps(k^2) &=  \left(
\begin{array}{cc}
 P^\eps_{hh}(k) & P^\eps_{hm}(k) \\
 P^\eps_{hm}(k) & P^\eps_{mm}(k) \\
\end{array}\right)\,,
\ea
and $\dagger$ denotes the transpose and complex conjugate.
We can now integrate out the stochastic fields in \refeq{epsdef} to \emph{derive} a likelihood for $\d_h(\vk)$ given the known analytic scaling of the noise-field correlators on large scales:
\ba
P\Big(\d_h(\vk) - \d_{h,\rm det}(\vk)\Big) &= \int d\eps_m(\vk)\:
P\Big(\d_h(\vk) - \d_{h,\rm det}(\vk) - b_1 \eps_m(\vk),\  \eps_m(\vk)\Big) \vs
&= (2\pi)^{-1/2} \left|  P^\eps_{hh}(k) + 2 b_1 P^\eps_{hm}(k) + b_1^2 P^\eps_{mm} \right|^{-1/2} \vs
& \qquad\times \exp\left[ -\frac12 \frac{\left|\d_h(\vk) - \d_{h,\rm det}(\vk)\right|^2 }{ P^\eps_{hh}(k) + 2 b_1 P^\eps_{hm}(k) + b_1^2 P^\eps_{mm} } \right]\,.
\label{eq:PcondFT1}
\ea
Given the diagonal covariance of the noise fields in Fourier space, we can then multiply the probabilities for the different wavenumbers $\vk$, leading to the following conditional probability for the halo field in Fourier space, up to an irrelevant normalization constant:
\ba
- \ln P\left(\vec{\d}_h\Big|\vec{\d}, \{b_O\}, \{\lambda_a\}\right) &=
\int_{\vk}^{k_{\rm max}}
\left[ \frac12 \ln \bar{\s}^2(k)
+  \frac1{2\bar{\s}^2(k)}
\left|\d_h(\vk) - \d_{h,\rm det}[\vec{\d},\{b_O\}](\vk)\right|^2 \right]\,,
\label{eq:PcondGFTc}
\ea
where the scale-dependent variance is given by
\ba
\bar{\s}^2(k) &=  P^\eps_{hh}(k) + 2 b_1 P^\eps_{hm}(k) + b_1^2 P^\eps_{mm}(k) \vs
&=   P^{\eps,0}_{hh} + \left( P^{\eps,2}_{hh} + 2 b_1 P^{\eps,2}_{hm} \right) k^2
  + \left( P^{\eps,4}_{hh} + 2 b_1 P^{\eps,4}_{hm} + b_1^2 P^{\eps,4}_{mm} \right) k^4
\,.
\label{eq:sigmabark}
\ea
Here, we have included terms up to order $k^4$, which are higher order, but ensure a positive variance. The integral in \refeq{PcondGFTc} is limited to wavenumbers up to some maximum value $k_{\rm max}$, which is a meta-parameter of the likelihood that can be varied. This cutoff ensures that the assumptions on the Gaussianity and analytic power spectra of the stochastic fields are fulfilled if $k_{\rm max}$ is chosen to be sufficiently small. From the EFT standpoint, \refeq{PcondGFTc} is thus the well-defined, unique likelihood for the large-scale biased-tracer field $\d_h(\vk)$ at the field level. This derivation is based on two key EFT results: the fact that on the scales of interest, there is a finite number of fields $O$ which describe the dependence of $\d_h$ on large-scale perturbations; and that the residual is approximately Gaussian-distributed with a power spectrum of known form (analytic in $k$). The scales controlling the validity of the bias expansion are well known (see Sec. 4.1.3 of \cite{biasreview} for a discussion). The scale controlling the Gaussianity of the likelihood has not been investigated in detail so far. We turn to that in \refsec{nongauss}.

A possible concern with a Fourier-space likelihood is that effects present in real world surveys such as masks are difficult to incorporate. Briefly, a mask would be incorporated by multiplying both $\d_{h,\rm det}$ and $\eps_h$ with the mask in real space. This leads to a non-diagonal covariance in \refeq{PcondGFTc} which is given by a convolution of white noise with the mask. Fortunately, since the mask is fixed, the computation and inversion of this covariance has to be performed only once. In the following, we will continue to assume a trivial mask, as appropriate for applications to simulations with periodic boundary conditions.

Let us now consider a finite volume, in particular a cubic box of side length $\Lbox$, and move to box normalization for the Fourier-space fields [\refeq{boxnorm}]. \refeq{PcondGFTc} immediately becomes
\ba
\ln P\left(\vec{\d}_h\Big|\vec{\d}, \{b_O\}, \{\lambda_a\}\right) &=
-\fsum{\vk}
\left[ \frac12 \ln \s^2(k)
+  \frac1{2\s^2(k)}
\left|\d_h(\vk) - \d_{h,\rm det}[\vec{\d},\{b_O\}](\vk)\right|^2 \right]\,,
\label{eq:PcondGFT}
\ea
where $\s^2(k)$ is now a dimensionless variance given by
\ba
\s^2(k) &\equiv \Lbox^{-3} \bar{\s}^2(k)
= V^{\eps,0}_{hh} + \left( V^{\eps,2}_{hh} + 2 b_1 V^{\eps,2}_{hm} \right) k^2
  + \left( V^{\eps,4}_{hh} + 2 b_1 V^{\eps,4}_{hm} + b_1^2 V^{\eps,4}_{mm} \right) k^4
\,,
\label{eq:sigmak}
\ea
and we have defined the noise variance parameters
\be
V^{\eps,n}_{xy} \equiv \Lbox^{-3} P^{\eps,n}_{xy}\,.
\ee

It is worth pointing out the very different interpretation of the EFT conditional probability derived here as compared to standard forward modeling approaches which use a local, real-space conditional probability as in \refeq{Pcond}. Instead of approximating a likelihood of unknown shape that is localized in physical space, we are expanding the \emph{perturbatively known non-local Fourier-space likelihood} up to a cutoff $k_{\rm max}$. This point will become more clear in the following. Note that one can generalize the diagonal covariance to a non-diagonal one, including the non-Gaussian contributions generated by nonlinear evolution. Again, we will turn to this in \refsec{nongauss}.

Finally, as long as errors in the forward model are captured by the likelihood \refeq{PcondGFT}, the full posterior \refeq{post} for the initial phases, cosmological parameters, and bias parameters becomes
\ba
P\left(\vec{\d}_\text{in},\theta,\{b_O\},\{\lambda_a\}\Big| \vec{\d}_h\right) =\:&
\mathcal{N}_P' P_\text{prior}(\vec{\d}_\text{in}|\theta) 
\exp \left[ \ln P\left(\vec{\d}_h\Big|\vec{\d}_\text{fwd}[\d_\text{in},\theta], \{b_O\},\{\lambda_a\}\right) \right]\,,
 \label{eq:postF}
\ea
where
\ba
\{ b_O \} &= \{ b_1 \cs  b_2 \cs b_{K^2} \} \cup \{ \clapl \} \vs
\{\lambda_a \} &= \{ V^{\eps,0}_{hh} \cs V^{\eps,2}_{hh} \cs V^{\eps,2}_{hm} \cs
V^{\eps,4}_{hh} \cs V^{\eps,4}_{hm} \cs V^{\eps,4}_{mm} \}\,.
\label{eq:biasfid}
 \ea
 In practice, we can limit the set $\{\lambda_a\}$ to the first three parameters, as the others are higher order (see the next section and \refapp{sigmak}).

\subsection{Maximum-likelihood point for bias parameters}

\refeq{postF} is clearly a highly complex, nonlinear and nonlocal (in terms of $\vec{\d}_\text{in}$) posterior. In order to make progress in our physical understanding, let us assume that we have fixed the initial density field $\vec{\d}_\text{in}$ to the true field. This thought example is easy to realize when applying the forward model to the results of simulations, whose initial conditions are known. For now, we will also fix the cosmology parameters $\theta$ to their true values.

Let us then consider the maximum-likelihood point for the bias parameter $b_O$, keeping all other parameters fixed (we will turn to the parameters controlling the variance below). For a fixed operator $O$, the maximum-likelihood point is given by:
\ba
-\frac{\partial}{\partial b_O} \ln P\left(\vec{\d}_h|\vec{\d}, \{b_O\}, \{\lambda_a\}\right) =\:& \fsum{\vk} \frac1{\s^2(k)}
O(\vk) \left(\d_h - \d_{h,\rm det}[\vec{\d},\{b_O\}]\right)^*_{\vk} \shouldeq 0 \,.
\ea
In the limit of infinite volume (e.g., a large number of simulation realizations), the products of fields in this relation approach their ensemble averages, i.e. their cross-power spectra. Then, the maximum-likelihood point becomes
\ba
\sum_{\vk}^{k_{\rm max}} \frac1{\s^2(k)}\< O(\vk) \d_h^*(\vk)\>  =\:& \sum_{O'} \hat b_{O'} \fsum{\vk} \frac1{\s^2(k)} \< O(\vk) O^\prime{}^*(\vk)\> \qquad\forall\  O \vs
& \stackrel{O=\d}{+}  V^{\eps,2}_{hm} k^2 + V^{\eps,4}_{hm} k^4 + b_1 V^{\eps,4}_{mm} k^4
\,. \label{eq:maxlikeGFourier}
\ea
The last line, which comes from the $b_1$-dependence of $\s^2(k)$ [\refeq{sigmak}] is only present when considering the operator $\d$. 
\refeq{maxlikeGFourier} corresponds to matching specific filtered (and weighted) moments in Fourier space, with a correction that scales analytically with $k$ and takes into account the correlation of noise in the halo field and matter.

By restricting $k_{\rm max} < k_{\rm NL}$, the nonlinear scale, these filtered moments can be kept under perturbative control. Recall that the operators $O$ here are still constructed by taking nonlinear transformations of the density field on the grid. Thus, there are now two filters involved: the kernel corresponding to the density assignment, and the sharp-$k$ filter for the moments on the larger length scale $1/k_{\rm max}$. This is in fact closely related to the approach followed by \cite{lazeyras/schmidt:2018,abidi/baldauf:2018}, who showed that one can efficiently obtain bias parameters up to cubic order using this approach.

Finally, note that including higher-derivative contributions in the bias expansion \refeq{dhdet} is very simple in the EFT likelihood, by generalizing $\d_{h,\rm det}$ in Fourier space to
  \be
\d_{h,\rm det}[\vec{\d},\{b_O\}](\vk) = \sum_O b_O O(\vk) \longrightarrow \sum_O \left[b_O - c_{\lapl O} k^2 \right] O(\vk)\,,
  \ee
  introducing an additional effective higher-derivative bias parameter $c_{\lapl O}$ which can be marginalized over to take into account improperly-modeled higher-order contributions. We continue to only include such a higher-derivative contribution $\propto \clapl$ for the density field, $O=\d$. We will see in \refsec{PT} why this is sufficient.

In order to gain a more explicit understanding of \refeq{maxlikeGFourier}, let us drop the second line, present if $O = \d$, for the remainder of this section, and continue to take the infinite-volume limit. We then have
\ba
\fsum{\vk} \frac1{\s^2(k)} \< O(\vk) \d_h^*(\vk) \>  = \fsum{\vk} \frac1{\s^2(k)} \sum_{O'} \hat b_{O'} \< O(\vk) O'^*(\vk)\> \qquad\forall\  O\,,
\label{eq:maxlikeG}
\ea
which is to be solved for the bias parameters $\hat b_O$. 
Let us consider this equality at fixed $\vk$.
If the set of bias operators in \refeq{dhdet} is of size $N_O$, \refeq{maxlikeG} is a linear system of $N_O$ equations that can be straightforwardly solved, yielding an estimator $\hat{\v{b}}(\vk)$ for the set of coefficients (see also \cite{lazeyras/schmidt:2018,abidi/baldauf:2018}):
\ba
\hat{\v{b}}(\vk) = \v{M}^{-1}(\vk) \cdot \v{H}(\vk)\,, \qquad \v{M} = \{ \< O O'^* \> \}_{O,O'}\,;\quad \v{H} = \{ \< \d_h^* O \> \}_O\,.
\label{eq:maxlikebO}
\ea
\refeq{maxlikeG} then corresponds to a weighted mean of the estimated bias parameters $\hat{\v{b}}(\vk)$ over $\vk$.

Similarly, taking the derivative of \refeq{PcondGFT} with respect to $\s^2(k)$ yields maximum-likelihood values for the various components of $\sigma^2(k)$, as derived in \refapp{sigmak}. In particular, the ML point for the constant part $V^{\eps,0}_{hh} \equiv \lim_{k\to 0} \sigma^2(k)$ is given by, in the infinite-volume limit:
\be
0 \shouldeq \fsum{\vk}\frac1{\s^4(k)} \left[\frac12 \s^{2}(k) -
  \< \left|\d_h(\vk) - \d_{h,\rm det}(\vk)\right|^2 \> \right]\,.
\label{eq:maxlikesigma2}
\ee
Thus, inaccuracies in the model directly contribute to the effective variance in the likelihood (assuming one allows $\s^2$ to vary). A deficient model thus lowers the amount of information that can be extracted, as expected. In order for the likelihood to be consistent however, it is of course necessary that the noise correlator $\< |\d_h-\d_{h,\rm det}|^2(\vk) \>$ is in fact analytic in $k$, as assumed in \refeq{sigmak}. We will see in the next section why and under what conditions this holds. At the order we work in throughout this paper, it is sufficient to include terms up to order $k^2$ in $\sigma^2(k)$, which corresponds to the first three parameters in \refeq{biasfid}. While the first, $V_{hh}^{\eps,0}$ corresponds to the physical halo shot noise in the large-scale limit, the other two contributions are effective stochastic parameters which absorb residuals of the model.\footnote{Indeed, beyond the large-scale limit there exists no unique definition of stochasticity (e.g., \cite{baldauf/schaan/zaldarriaga,baldauf/schaan/zaldarriaga:2}).}

\subsection{Including cosmological parameters}
\label{sec:sigma8}

We now move on to the estimation of cosmological parameters. The most simple, and interesting, cosmological parameter in this context is the normalization of the primordial perturbations (or equivalently linear matter power spectrum). Following convention, we parametrize this through the variance $(\s_8)^2$ at redshift zero of the linear density field filtered with a real-space tophat kernel on the scale $8\Mpch$. This parameter is interesting, since, at the linear level, it is exactly degenerate with the linear bias $b_1$. Thus, we need nonlinear information in the measured density field $\vec{\d}_h$ to break this degeneracy, which is a highly nontrivial test of the forward model. We will discuss other cosmological parameters in \refsec{concl}.

In order to investigate this, we assume that the operators appearing in \refeq{dhdet} [e.g., \refeq{Osetfid}] are split by perturbative order so that each scales homogeneously with $\s_8$. For convenience, we define $\alpha \equiv \s_8/\s_{8,\rm fid}$, where the fiducial value $\s_{8,\rm fid}$ is fixed. We can then write
\be
O(\sigma_8) = \alpha^{n_{s,O}} O(\s_{8,\rm fid})\,,
\ee
so that, e.g., $n_{s,\d^{(1)}} = 1, n_{s,\d^{(2)}} = n_{s,\d^2} = 2,$ and so on. We then have
\be
\vec{\d}_{h,\rm det}(\{b_O\},\s_8) = \sum_O b_O \alpha^{n_{s,O}} \vec{O}(\s_{8,\rm fid})\,.
\ee
In the following, all operators $O$ will be assumed to be evaluated at $\s_{8,\rm fid}$, and we will omit this dependence for clarity. We obtain, again neglecting the terms in the second line of \refeq{maxlikeGFourier},
\ba
-\frac{\partial}{\partial b_O} \ln P\left(\vec{\d}_h|\vec{\d}, \{b_O\}, \{\lambda_a\}\right) =\:& \fsum{\vk} \frac1{\s^2(k)}
\alpha^{n_{s,O}} O(\vk) \left(\d_h - \d_{h,\rm det}[\vec{\d},\{b_O\}]\right)^*_{\vk} \shouldeq 0 \label{eq:maxlikebs-1}\\
-\frac{\partial}{\partial\ln \s_8} \ln P\left(\vec{\d}_h|\vec{\d}, \{b_O\}, \{\lambda_a\}\right) =\:& \sum_O b_O n_{s,O} \alpha^{n_{s,O}} \fsum{\vk} \frac1{\s^2(k)}
O(\vk) \left(\d_h - \d_{h,\rm det}[\vec{\d},\{b_O\},\alpha]\right)^*_{\vk} \shouldeq 0\,.
\nonumber
\ea
In the infinite-volume limit and at fixed $\vk$, we thus have
\ba
\< O \d_h^* \>_{\vk} \shouldeq\:&
 \sum_{O'} b_{O'} \alpha^{n_{s,O'}} \< O O'^* \>_{\vk} \label{eq:maxlikebs0} \\
 \sum_O b_O n_{s,O} \alpha^{n_{s,O}} \< O \d_h^* \>_{\vk} \shouldeq\:&
\sum_{O,O'} b_O n_{s,O} b_{O'} \alpha^{n_{s,O}+n_{s,O'}} \< O O'^* \>_{\vk}\,.
\nonumber
\ea
We now define scaled parameters
\be
\beta_O \equiv b_O \alpha^{n_{s,O}}\,.
\ee
\refeq{maxlikebs0} then becomes
\ba
\< O \d_h^* \>_{\vk} \shouldeq\:&
 \sum_{O'} \beta_{O'}  \< O O'^* \>_{\vk} \vs
 \sum_O n_{s,O} \beta_O \< O \d_h^* \>_{\vk} \shouldeq\:&
 \sum_{O,O'}  n_{s,O} \beta_O \beta_{O'}  \< O O'^* \>_{\vk}\,.
 \label{eq:maxlikebs1}
\ea
Contracting the first line with $\sum_O n_{s,O}\beta_O$ immediately shows that the second line is linearly dependendent on the first line, which contains $N_O$ equations. That is, \refeq{maxlikebs1} in fact contains (at most) $N_O$ independent equations for $N_O+1$ unknowns. This is not surprising: if all operators have independent bias parameters, then any change in $\s_8$ can be absorbed by a change in the $N_O$ bias parameters. Thus, we need to rely on some relations between bias parameters of different operators. In our fiducial application of the EFT likelihood, these are $\d^{(1)}$ and $\d^{(2)}$ which are both multiplied by $b_1$.

At fixed $\vk$, we thus have for the maximum-likelihood point of $\v{\beta}$:
\ba
\v{H}(\vk) &\shouldeq \v{M}(\vk) \cdot \v{\beta}(\vk) \vs
\Rightarrow\quad
\hat{\v{\beta}}(\vk) &= \v{M}^{-1}(\vk) \cdot \v{H}(\vk)\,.
 \label{eq:maxlikebs2}
\ea
This relation is of course of the same form as derived above in \refeq{maxlikebO} (without considering $\s_8$). Crucially, in order to break the degeneracy between bias parameters and $\s_8$, there must be at least two non-degenerate operators multiplied by the same bias parameters, which scale differently with $\s_8$. Physically, this is the case at second order in perturbations due to the displacement term inside $\d^{(2)}$. Since large-scale galaxy displacements have to be the same as those of matter due to the equivalence principle (see Sec.~2.7 of \cite{biasreview} for a discussion), this displacement term has to be multiplied by the same bias $b_1$ as the linear density field. The estimation of $\sigma_8$ using the EFT likelihood is hence based only on this fundamental physical constraint. 

Finally, the full maximum-likelihood point then corresponds to a weighted average over $\vk$, following \refeq{maxlikebs-1} [and, in practice, including the contributions in the second line of \refeq{maxlikeGFourier}].

\section{The maximum-likelihood point of the EFT likelihood}
\label{sec:PT}

In this section, we will derive predictions in the EFT for the correlators $\<\d_h O\>$ and $\<O O'\>$ which enter in the maximum-likelihood point \refeq{maxlikeGFourier}. This allows us to connect inferences based on the EFT likelihood to well-known predictions for matter and halo $n$-point functions.
As we will see, it is possible to consistently remove the dependence of the predictions on non-perturbative small-scale modes by renormalization.
Throughout, we will work up to cubic order in perturbation theory, which essentially means that all contributions can be expressed as linear or quadratic functionals of the linear matter power spectrum $\Plin(k)$. We also include the leading higher-derivative contributions which scale as $k^2\Plin(k)$. Higher-order terms which we neglect throughout involve three power spectra as well as higher powers of $k^2$. We discuss these in \refsec{PT:HO}.

We will adopt the fiducial operator set described in \refsec{bias}, which involves the complete set of bias operators up to second order:
\ba
O \in \Oset &= \Oset^{[1]} \cup \Oset^{[2]} \vs
\Oset^{[1]} &= \{ \d,\  \lapl\d \}\,,\quad \Oset^{[2]} = \{ \d^2,\  K^2 \}\,,
\ea
with coefficients
\be
\{ b_1,\  \clapl,\  b_{\d^2},\  b_{K^2} \}
= \{ b_1,\  \clapl,\  (b_2/2),\  b_{K^2} \}\,.
\ee
We have not included cubic-order bias terms. Only one cubic bias term is relevant at the level of the next-to-leading order power spectrum and leading-order three-point function (specifically, $b_{\otd}$ in \refeq{PhmNLO}; see also Sec.~4.1 of \cite{biasreview}), which are the relevant statistics here. Moreover, at the order we work in, its contribution is almost perfectly degenerate with that of the higher-derivative operator $\lapl\d$. As we will show, the maximum-likelihood point yields the physical values of $b_1\cs b_2 \cs b_{K^2}$, while $\clapl$ is an effective, smoothing scale-dependent parameter which absorbs the cubic bias contribution as well.

We will now re-introduce the subscript $W$ for filtered fields.
Unless otherwise noted, we allow for a general filter function $W(\vx)$ in real space, with Fourier-space version $W(\vp)$ which satisfies $W(-\vp) = W^*(\vp)$. As we will see, the requirement of unbiased results from the EFT likelihood places constraints on the form of the filter.

\subsection{Summary}
\label{sec:main_summary}

Since the remainder of this section is a somewhat long and technical discussion, let us summarize the findings here. The key result is that \refeq{maxlikeGFourier} indeed yields \emph{unbiased results for $\s_8$ and the bias parameters which are under perturbative control}, if the following conditions are met:
  \begin{enumerate}
  \item A sharp-$k$ filter $W_\L(\vk)$ [\refeq{sharpk}], where $\L$ is a cutoff scale, is used to obtain the smoothed density field out of which the operators $O$ are constructed.
  \item $k_{\rm max}$ is smaller than the cutoff $\L$ of this filter.
  \item The operators $O \to [O]$ are constructed as \emph{renormalized} operators, as explained in \refsec{renorm}, where counterterms that are relevant at the perturbative order we work in are subtracted.
  \item The list of operators includes $\lapl \d(\vx) \leftrightarrow -k^2 \d(\vk)$, and the likelihood involves a specific scale-dependent variance term $\s^2(k)$ which scales analytically with $k$ [\refeq{sigmak}] and depends on $b_1$. Both of these ingredients are important to consistently absorb higher-order contributions.
  \end{enumerate}
Readers not interested in the technical details of how these results are obtained can skip to the next section.

\subsection{Renormalization conditions}
\label{sec:renorm}

When including nonlinear operators in the bias expansion, such as density squared $\d^2$, in the context of the general perturbative bias expansion, it is important to construct these operators in such a way that their cross-correlations are not sensitive to small-scale perturbations. Technically, this is achieved by employing \emph{renormalized operators} $[O]$, which are obtained from the \emph{bare} operators $O$ by subtracting counterterms.

We follow the general approach by Ref.~\cite{assassi/etal}, who derived the following renormalization conditions:
\be
 \< [O](\vk) \d^{(1)}(\vk_1) \cdots \d^{(1)}(\vk_n)\>
\stackrel{\{k_i\} \rightarrow 0}{\simeq}
  \< O(\vk) \d^{(1)}(\vk_1) \cdots \d^{(1)}(\vk_n) \>_{\LO}\,,
\label{eq:renormcond}
\ee
where $n=0,1,2,...$. Here, $\<\cdot\>_{\LO}$ stands for a correlator evaluated at leading order (LO) in perturbation theory. \refeq{renormcond} ensures that all large-scale cross-correlations between the renormalized operators $[O]$ are independent of small-scale modes.

The $n=0$ condition simply reads $\< [O](\vk) \> = 0$, which we have already enforced in \refeq{Osetfid}. Beyond this, at the perturbative order which we work in, it is sufficient to only consider the conditions for $n=1$ for the quadratic operators $\d^2$ and $K^2$.\footnote{Here, there is a subtlety if the quadratic operators are constructed from sharp-$k$ filtered fields. This is discussed in \refapp{renorm}.}
As shown in \refapp{renorm}, the renormalization conditions can be enforced by subtracting counterterms in Fourier space as follows:
\ba
[\d](\vk) &= \d(\vk) \vs
\Bigl[ \d^2 \Bigr] &= (\d^2)(\vk) - \Sigma_{1-3}^2(k) \d(\vk)
\quad\mbox{and}\quad [\d^2](\vk=0) = 0 \vs
\Bigl[K^2 \Bigr](\vk) &= (K^2)(\vk) - \frac23 \Sigma_{1-3}^2(k) \d(\vk)
\quad\mbox{and}\quad [K^2](\vk=0) = 0\,,
\label{eq:renormfid}
\ea
where all fields are presumed filtered with $W$, and $\Sigma_{1-3}^2(k)$ is defined in \refeq{Sigma13def}. The conditions at $\vk=0$ are given for completeness here, even though they do not matter in practice, as the sum in the likelihood excludes the zero mode [\refeq{PcondGFT}].
The density field $\d$ itself does not need to be renormalized (up to higher-derivative terms), as it is protected by mass and momentum conservation.
The reason that we subtract a $k$-dependent counterterm, rather than the asymptote in the limit $k\to 0$ [given in \refeq{Sigma130}] is that, otherwise, residuals of order $k/\Lambda$ are present which are not strictly absorbed by the effective higher-derivative bias $\clapl$.

The counterterms in \refeq{renormfid} are simply proportional to the density field $\d$. One might wonder why they then cannot be absorbed in the bias coefficient $b_1$. The reason is that, as explained after \refeq{maxlikebs2}, a correct inference of cosmology using the forward model relies on a consistent treatment of bias, in particular the fact that the same bias parameter $b_1$ multiplies both the linear and second-order density fields. This consistency is broken if the quadratic operators are not renormalized following \refeq{renormfid}. Moreover, \refeq{renormfid} also removes the leading large-scale connected contribution (trispectrum) to the operator cross-correlations on the r.h.s. of \refeq{maxlikeGFourier}, as shown in \refapp{renorm}.

\subsection{Conditions for unbiased maximum-likelihood point}

We now explicitly derive the expressions obtained on the left-hand side and right-hand side of the maximum-likelihood point of the Gaussian likelihood in Fourier space,
\ba
\fsum{\vk} \frac1{\s^2(k)}\< [O](\vk) \d_h^*(\vk)\>  =\:& \sum_{O'} \hat b_{O'} \fsum{\vk} \frac1{\s^2(k)} \< [O](\vk) [O'{}^*](\vk)\> \qquad\forall\  O \vs
& \stackrel{O=\d}{+} V^{\eps,2}_{hm} k^2 + V^{\eps,4}_{hm} k^4 + b_1 V^{\eps,4}_{mm} k^4
\,,
\label{eq:maxlikeGFourier2}
\ea
at leading order in the general perturbative bias expansion (or equivalently, the effective field theory of biased tracers). As we will see, this involves explicit results at second order in perturbation theory, and a subset of third-order contributions. Moreover, it is sufficient to focus on a single unspecified value of $\vk$ here (with $|\vk| < k_\text{max}$), since \refeq{maxlikeGFourier2} corresponds to a weighted sum over $\vk$. Finally, for calculational simplicity, we return to the continuum limit to calculate the correlators in this relation. 
We give the summary of results here, with detailed calculations relegated to \refapp{opcorr}. In the following, we will make the smoothing filter $W$ explicit again, as it is of crucial relevance in this derivation.

\subsubsection{Quadratic operators}
\label{sec:PT:MLEO2}

Let us begin with \refeq{maxlikeGFourier2} evaluated for a quadratic operator $O = O^{[2]}$. Then, the second line is not present.
Any \emph{bare} quadratic operator can be expressed in Fourier space as
\be
O^{[2]}[\d_W](\vk) = \int_{\vp} S_O(\vp,\vk-\vp) \d(\vp) \d(\vk-\vp) W(\vp) W(\vk-\vp)\,,
\ee
where $S_O$ is a kernel listed for the relevant operators in \reftab{kernels}. We will neglect the counterterms for now, and include them at the end of this section. Thus, the Fourier-space correlator $\< \d_{h,W}(\vk) O^{[2]}[\d_W](\vk') \>'$ corresponds to an integral over the halo-matter-matter bispectrum:
\ba
& \< \d_{h,W}(\vk') O^{[2]}[\d_W](\vk) \>' = W(\vk') \int_{\vp} S_O(\vp,\vk-\vp) W(\vp) W(\vk-\vp) \< \d_h(\vk') \d(\vp) \d(\vk-\vp) \>' \vs
&\quad =  W(-\vk) \left[b_1 - b_{\lapl\d} k^2 \right]\int_{\vp} S_O(\vp,\vk-\vp) W(\vp) W(\vk-\vp) B_{mmm}(-\vk, \vp, \vk-\vp) \vs
&\quad\  + 2 W(-\vk) \sum_{O' \in \Oset^{[2]}} b_{O'} \int_{\vp} S_{O'} (\vp,\vk-\vp) S_O(\vp,\vk-\vp) W(\vp) W(\vk-\vp) \Plin(p)\Plin(|\vk-\vp|)\,,
\ea
where a prime on an expectation value denotes that the momentum-conserving Dirac delta is to be dropped. In the second line we have used $\vk'=-\vk$ from momentum conservation as well as the result for the tree-level halo-matter-matter bispectrum and the tree-level matter bispectrum, given in \refapp{PTref}.

The operator correlations appearing on the right-hand side of \refeq{maxlikeGFourier2} are given by
\ba
\<  O^{[2]}[\d_W](\vk) \d_W(\vk') \>' =\:& W(-\vk) \int_{\vp} S_O(\vp,\vk-\vp) W(\vp) W(\vk-\vp) B_{mmm}(-\vk, \vp, \vk-\vp) \vs
\<  O^{[2]}[\d_W](\vk) O^{[2]}{}'[\d_W](\vk') \>' =\:& 2 \int_{\vp} S_O(\vp,\vk-\vp) S_{O'}(\vp,\vk-\vp) |W(\vp)|^2 |W(\vk-\vp)|^2 \vs
& \hspace*{3cm}\times \Plin(p) \Plin(|\vk-\vp|)\,.
\label{eq:O2O2bare}
\ea
We can now evaluate the equality corresponding to the maximum-likelihood point of the Fourier-space likelihood, \refeq{maxlikeGFourier}. For clarity, we restrict to a fixed value of $\vk$. We then obtain, for any $O \in \Oset^{[2]}$,
\ba
& W(-\vk) \left[b_1 - b_{\lapl\d} k^2 \right] \int_{\vp} S_O(\vp,\vk-\vp) W(\vp) W(\vk-\vp) B_{mmm}(-\vk, \vp, \vk-\vp)
\label{eq:O2lhsrhs}\\
&\qquad + 2 W(-\vk) \sum_{O' \in \Oset^{[2]}} b_{O'} \int_{\vp} S_{O'} (\vp,\vk-\vp) S_O(\vp,\vk-\vp) W(\vp) W(\vk-\vp)
\Plin(p)\Plin(|\vk-\vp|) \vs
& \shouldeq
b_1 W(-\vk) \int_{\vp} S_O(\vp,\vk-\vp) W(\vp) W(\vk-\vp) B_{mmm}(-\vk, \vp, \vk-\vp) \vs
& \qquad\qquad + 2 \sum_{O' \in \Oset^{[2]}} b_{O'} \int_{\vp} S_O(\vp,\vk-\vp) S_{O'}(\vp,\vk-\vp) |W(\vp)|^2 |W(\vk-\vp)|^2
\Plin(p) \Plin(|\vk-\vp|)\,.
\nonumber
\ea
First, note that the term involving $b_{\lapl\d}$ involves a higher power of $k$ than the others; that is, it is higher order in the perturbative counting. We consider such higher-order contributions in \refsec{PT:HO}.

Disregarding this term, we see from \refeq{O2lhsrhs} that, already at leading order, the equality \refeq{maxlikeGFourier2} only yields unbiased results if $(i)$ the smoothing filter $W(-\vk)$ is unity at the scales $|\vk| < k_{\rm max}$ considered; $(ii)$ the smoothing filter satisfies
\be
W(\vk) = |W(\vk)|^2\,.
\ee
This corresponds to a \emph{sharp-$k$} filter of the form
\be
W(\vk) = W_\Lambda(\vk) \equiv \thH( \Lambda - \norm{\vk})\,,
\label{eq:sharpk}
\ee
where $\norm{\cdot}$ is a norm, e.g. $\sqrt{k_i k^i}$ (spherical tophat) or $\sum_i |k_i|$ (cube).
One can show that the same conclusion holds if one were to use a real-space likelihood that corresponds to matching correlators at \emph{finite lag}. That is, the result on the filter shape is not simply a coincidence caused by our working with a Fourier-space likelihood. Clearly, for such a sharp-$k$ filter function, $W_\L(\vk) = 1$ holds as long as $k_{\rm max} < \L$, so that condition $(i)$ is satisfied trivially.

Finally, let us turn to the counterterms which we have not included so far. The contribution to $\< \d_{h,W} [O^{[2]}]\>$ on the left-hand side (l.h.s.) of \refeq{maxlikeGFourier} cancels with that to $b_1 \< \d_W [O^{[2]}]\>$ on the right-hand side (r.h.s.). On the other hand, as shown in \refapp{renorm}, the counterterms precisely remove the leading connected trispectrum contribution to $\< O^{[2]} O'^{[2]}\>$, which is proportional to $\< \d_W^2\>^2 \Plin(k)$ and which we have not written in \refeq{O2O2bare}, and thus ensure that the equality corresponding to the maximum-likelihood point holds at the relevant order.

\begin{table*}[b]
\centering
\begin{tabular}{l|l}
\hline
\hline
Operator & Kernel $S_O(\vk_1,\vk_2)$ \\
\hline
$\d^{(2)}$ & $F_2(\vk_1,\vk_2)$ \\
$\d^2$  & 1 \\
$K^2$ & $\mu_{12}^2 - 1/3$ \\
$s^k\partial_k \d$ & $-\mu_{12} (k_1/k_2 + k_2/k_1)/2$ \\
\hline
\hline
\end{tabular}
\caption{Fourier-space kernels $S_O(\vk_1,\vk_2)$ corresponding to the quadratic operators appearing in the halo and matter $n$-point functions. We have denoted $\mu_{12}\equiv\vkhat_1\cdot\vkhat_2$, and the kernel $F_2$ is given in \refeq{F2}.}
\label{tab:kernels}
\end{table*}

\subsubsection{Density}
\label{sec:PT:MLEd}

Let us now turn to the operator $O[\d_W] = \d_W$ in $\Oset^{[1]}$. On the left-hand side of \refeq{maxlikeGFourier2}, we have
\be
\< \d_{h,W}(\vk')\  [\d_W](\vk) \>' \equiv \< \d_{h,W}(\vk') \d_W(\vk) \>' = |W(\vk)|^2 P_{hm}(k) =
|W(\vk)|^2 \left[b_1 \Plin(k) + P_{hm}^\NLO(k)\right]\,,
\nonumber
\ee
where $P_{hm}^\NLO(k)$ is the next-to-leading order (NLO) contribution to the halo-matter power spectrum, which we can write as (see \refapp{PTref})
\be
P_{hm}^\NLO(k) = b_1 P_{mm}^\NLO(k) + \hat P_{hm}^\NLO(k)\,,
\ee
where $P_{mm}^\NLO(k)$ is the NLO contribution to the matter power spectrum, and $\hat P_{hm}^\NLO(k)$ includes the nonlinear bias terms. We include the NLO contributions to the power spectra here, since they are of the same order as other terms that will appear on the right-hand side.

The operator correlations appearing on the right-hand side of \refeq{maxlikeGFourier2} are given by
\ba
\< \d_W(\vk) \d_W(\vk') \>' &= |W(\vk)|^2 P_{mm}(k) = |W(\vk)|^2 \left[\Plin(k) + P_{mm}^\NLO \right]\,, \vs
\< \d_W(\vk') \Big[O^{[2]}[\d_W]\Big](\vk) \>' &= W(-\vk) \int_{\vp} S_O(\vp,\vk-\vp) W(\vp) W(\vk-\vp) B_{mmm}(-\vk, \vp, \vk-\vp) \vs
&\quad - r_O \Sigma_{1-3}^2(k) |W(\vk)|^2 P_{mm}(k)
+  V^{\eps,2}_{hm} k^2 \,,
\nonumber
\ea
where $r_{\d^2} = 1$, $r_{K^2} = 2/3$, and the contribution $\propto \clapl$ is trivially related to the first line. In the second line, we have restricted the stochastic contributions in the second line of \refeq{maxlikeGFourier2} to the leading term $\propto k^2$, as the terms scaling as $k^4$ are subleading.

We can now evaluate the equality corresponding to the maximum-likelihood point of the Fourier-space likelihood, \refeq{maxlikeGFourier2}, for $O=\d_W$. For clarity, we again restrict to a fixed value of $\vk$. We then obtain
\ba
&
|W(\vk)|^2 \left[b_1 \Plin(k) + P_{hm}^\NLO(k) \right]\shouldeq
|W(\vk)|^2 \left\{ b_1 \left[\Plin(k) + P_{mm}^\NLO\right] - \clapl k^2 \Plin(k) \right\} \vs
& \hspace*{1.5cm}+ \sum_{O'\in \Oset^{[2]}} b_{O'} \bigg[ W(-\vk) \int_{\vp} S_{O'}(\vp,\vk-\vp) W(\vp) W(\vk-\vp) B_{mmm}(-\vk, \vp, \vk-\vp) \vs
 & \hspace*{3.8cm} -  r_{O'} \Sigma_{1-3}^2(k) |W(\vk)|^2 P_{mm}(k) \bigg] +  V^{\eps,2}_{hm} k^2\,.
\label{eq:dlhsrhs}
\ea
The evaluation of this equality follows along the same lines as for the quadratic operators discussed above. However, it is somewhat more technical. We thus summarize the result here, and refer to \refapp{opcorr} for the details.
The left-hand side of \refeq{dlhsrhs} contains the halo-matter cross-power spectrum up to NLO, i.e. including 1-loop contributions. As shown in \refapp{opcorr},
\emph{if} we again specialize to a sharp-$k$ filter $W(\vk) = W_\L(\vk)$,
the right-hand side is able to match this prediction consistently up to corrections that scale as $(k/\Lambda)^2\Plin(k)$, $(k/\knl)^2\Plin(k)$, and $(k/\Lambda)^2$. The first two types of residuals can be absorbed by $\clapl$, while the last type of residual is absorbed by $V^{\eps,2}_{hm}$. Note that this implies that the latter two parameters are effective, $\Lambda$-dependent parameters which differ in general from the physical parameters describing the halo field.

\subsection{Suppression of higher-order terms}
\label{sec:PT:HO}

Let us now discuss the approximate magnitude of the terms we have neglected throughout. For this we assume that $k_{\rm max}$ is set to be proportional to $\L$, i.e. $k_{\rm max} = \epsilon \L$ with $\epsilon < 1$, so that $\L$ is the only relevant scale. Further, we continue to use the continuum normalization.
The most rigorous approach to estimate the relevance of higher-order terms is to include the full set of next-to-leading terms and repeat the entire inference. The resulting shift in parameters then corresponds to the systematic error made by neglecting these terms. This clearly goes beyond the scope of the current paper, and we here instead follow a simpler approach, by providing an estimate of the subleading terms in the maximum-likelihood relation.

It is further worth noticing that, for any given higher-order bias field, we only need to be concerned with the part that correlates with the lower-order fields. The remainder functions as a stochastic contribution which is captured by the (scale-dependent) variance in the likelihood.

First, consider the approximate scaling of the contributions to the maximum-likelihood conditions, i.e. at the correlator level, which we have kept. The fractional size of the NLO contribution to the LO ones is at most as large as
\be
\frac{\text{NLO}}{\text{LO}}:\quad
(\L R_*)^2 \,,\ \left(\frac{\L}{\knl}\right)^{3+n} \,,\  \frac{\L^2 V^{\eps,2}_{hm}}{\Plin(\L)}\,,
\ee
where $R_*$ is the spatial scale associated with the tracer (i.e. roughly of order the Lagrangian radius for halos, although it is expected to remain at least of order $\knl^{-1}$ even for low-mass halos), and $n \approx -1.7$ is the slope of the linear matter power spectrum around the nonlinear scale. The first of the NLO contributions is higher-derivative, while the second is the loop contribution which we have estimated using a power-law ansatz (see e.g. \cite{pajer/zaldarriaga}).
The last term is the scale-dependent noise, which, given the form of the linear matter power spectrum at scales $\L \gtrsim 0.02 \iMpch$ scales faster with $\L$ than the others. It corresponds to the cross-correlation of the effective noise in the matter field with the halo shot noise, in addition to effective contributions as derived in \refapp{opcorr}. Its magnitude is difficult to predict a priori. However, if one assumes that the noise in the halo field dominates over that in the matter density field, and that the scale dependence of the former is controlled by the spatial scale $R_*$, one expects that $|V^{\eps,2}_{hm}| \lesssim R_*^2 P^{\eps,0}_{hh} \sim R_*^2/\bar n$ where $\bar n$ is the mean halo density, and the second scaling follows from Poisson shot noise (see also \refsec{nongauss}). 
The effective contribution to $|V^{\eps,2}_{hm}|$ which absorbs residuals in the maximum-likelihood relations scales as $\sim \L^{-2} \knl^{-3}$, as shown in \refapp{opcorr}, and thus is typically smaller.

We can now consider the corresponding scaling of higher-order contributions.
The leading neglected contributions are expected to scale as
\be
\frac{\text{NNLO}}{\text{LO}}:\quad  (\L R_*)^4\,,\ (\L R_*)^2\left(\frac{\L}{\knl}\right)^{3+n}\,,\
\left(\frac{\L}{\knl}\right)^{2(3+n)}\,,\  \frac{\L^4 V^{\eps,4}}{\Plin(\L)}\,.
\nonumber
\ee
Clearly, these corrections can be made as small as required by the experimental accuracy by reducing $\L$ (and $k_{\rm max}$).

Alternatively, one can include higher-order and higher-derivative bias contributions in the likelihood. However, other aspects of the likelihood including noise terms (\refsec{PT:quadrnoise}) and non-Gaussianity of the likelihood (\refsec{nongauss}) need to be evaluated carefully in order to ensure that the higher-order likelihood remains consistent. Further, a consistent higher-order EFT likelihood requires more counterterms to the quadratic as well as higher-order operators in order to ensure that \refeq{renormcond} holds at the corresponding order.

\subsection{Noise fields for quadratic operators}
\label{sec:PT:quadrnoise}

With the understanding gained in this section, let us now go back to
the starting point of the derivation of the EFT likelihood, \refeq{epsdef}.
There, we assumed that the only noise in the model that needed to be considered
is due to the matter field itself. We begin by reviewing the reason we had
to include this noise. In the EFT, the matter density field on large scales
contains a noise term that captures the coupling of two non-perturbative small-scale modes to a large-scale mode \cite{abolhasani/mirbabayi/pajer:2016,baldauf/schaan/zaldarriaga,baldauf/schaan/zaldarriaga:2}. This noise in turn can correlate with the
noise in the halo density field, since the latter also depends on the
small-scale modes. The forward model for $\vec{\d}_\text{fwd}[\vec{\d}_\text{in}]$ used
in the likelihood might or might not correctly capture the
noise $\eps_m$; in either case, we have to allow for this noise correlation,
which leads to the term $b_1 V^{\eps,2}_{hm}$ in $\sigma(k)$ [\refeq{sigmak}].
In this section, we discuss potential noise contributions appearing in the nonlinear operators, as well as the case of a noise amplitude that depends on large-scale perturbations. The main conclusion of this section is that either type of contribution is not relevant at the order we work in.

Let us thus consider the higher-order operators. These are constructed out
of the filtered density field, which only includes modes with $k < \Lambda$ which are under perturbative control. A potential concern is that the square
$\d_\Lambda^2$ of a field $\d_\Lambda$ that contains noise $\eps_m$ will lead to a white-noise contribution which is relevant on large scales. Since the noise $\eps_m$ is by definition uncorrelated with large-scale modes $k < \Lambda$, its contribution to the correlator of the bare operator $\d_\Lambda^2$ is given by
\ba
\< (\d_\Lambda)^2(\vk) (\d_\Lambda)^2(\vk')\>_{\eps_m}' =\:& \int_{\vp} \thH(\Lambda - p) \thH(\Lambda - |\vk-\vp|) P_{\eps_m}(p) P_{\eps_m}(|\vk-\vp|) \vs
\stackrel{k \to 0}{\simeq}\:& \int_{\vp} \thH(\Lambda-p) \left[P_{\eps_m}(p)\right]^2 + \O\left(\frac{k^2}{\Lambda^2}\right)\,.
\ea
First, note that we only need to worry about the zeroth order term in $k$ here, as the contribution $\propto k^2$ and higher powers can be absorbed by the scale-dependent noise terms in $\sigma^2(k)$ [\refeq{sigmak}].
Mass and momentum conservation require that $P_{\eps_m}(p) \propto p^4$. Let us thus roughly approximate
\be
 P_{\eps_m}(p) = C \frac{p^4}{\knl^7}\,.
 \ee
Ref.~\cite{baldauf/schaan/zaldarriaga}, who considered several different perturbative forward models for the density, showed that $C \lesssim 1$, although they did find some deviations from the $p^4$ scaling (note that $\knl$ is in the range $0.25-0.3\iMpch$ for a standard $\Lambda$CDM cosmology at $z=0$). This yields
\ba
\< (\d_\Lambda)^2(\vk) (\d_\Lambda)^2(\vk')\>_{\eps_m}'
\stackrel{k \to 0}{\simeq}\:&  \frac{C^2}{2\pi^2} \int_p^\Lambda dp
\frac{p^{10}}{\knl^{14}}
= \frac{C^2}{22\pi^2 \knl^3} \left(\frac{\Lambda}{\knl}\right)^{11}\,.
\ea
We see that this is very highly suppressed if $\Lambda < \knl$. However, even for $\Lambda=\knl$, the prefactor is quite small compared to the leading contribution to the maximum-likelihood equation for the quadratic operators considered in \refsec{PT:MLEO2}. Similar results hold of course for the operator $(K_\L)^2$. Thus, assuming that the noise in the forward model is not much larger than the residuals found by \cite{baldauf/schaan/zaldarriaga}, the contribution from $\eps_m$ to the correlators of the quadratic operators is negligible at the order we work in. Nevertheless, if necessary, a noise field $b_2 \eps_{\d^2}$ could be integrated out similarly to what is done for $b_1 \eps_m$ in \refsec{GaussF}.

In general, the amplitude of the noise can also depend on large-scale density fluctuations. This can be captured by writing, in real space,
\be
\eps_h(\vx) = \eps_{h,0}(\vx) + \eps_{h,\d}(\vx) \d_\L(\vx) + \cdots\,,
\label{eq:epshexp}
\ee
where the fields $\eps_{h,0}, \eps_{h,\d}, \cdots$ are again uncorrelated with the large-scale perturbations such as $\d_\L$ (see Sec.~2.8 of \cite{biasreview} for a detailed discussion). At the level of the power spectrum, the additional fields do not add any nontrivial contribution, as they can be absorbed into the other bias parameters. At the bispectrum level, the term $\eps_{h,\d}\d_\L$ in \refeq{epshexp} does contribute. However, it only appears in correlators that involve two powers of the halo density field, since the fields in $\eps_h$ only correlate with each other. As discussed in \refsec{PT:MLEO2}, only single powers of the halo density field appear in the maximum-likelihood equation of the Gaussian likelihood. Thus, the additional terms in \refeq{epshexp} are not relevant. They do in general contribute when going beyond the Gaussian likelihood, which we turn to in the next section.

\section{Beyond the Gaussian likelihood}
\label{sec:nongauss}

Let us now consider the limitations of the Gaussian likelihood derived
in \refsec{GaussF}. For this, we derive the expected size of the leading
non-Gaussian corrections in the EFT approach. We use the box normalization
in this section, since the precise counting of modes is relevant. Let us begin with the halo
noise field, and assume that it has a nonzero three-point function:
\ba
\< \eps_h(\v{n} k_F) \eps_h(\v{n}' k_F) \eps_h(\v{n}'' k_F) \> = \frac1{\Lbox^6} \d_{\v{n}+\v{n}',-\v{n}''}  B_{\eps_h}(\v{n} k_F, \v{n}' k_F) \,.
\ea
Strictly speaking, we need to consider the bivariate distribution for $\{ \vec{\eps}_h, \vec{\eps}_m \}$ and integrate out $\vec{\eps}_m$ following \refeq{PcondFT1}. However, we are mostly interested in the scaling of the leading correction to the Gaussian likelihood.
Let us thus use the Edgeworth expansion for $\eps_h$, so that the likelihood of the $\eps_h$ field including the leading non-Gaussian correction becomes (again up to an irrelevant normalization)
\ba
- 2 \ln P(\vec{\eps}_h) = \fsum{\vk} \Bigg[& \ln \sigma_{\eps_h}^2(k) + \frac{|\eps_h(\vk)|^2}{\sigma_{\eps_h}^2(k)} \vs
  & - \frac13 \fsum{\vk'} \frac{B_{\eps_h}(\vk,\vk')}{\Lbox^6 \sigma_{\eps_h}^2(k)\sigma_{\eps_h}^2(k')\sigma_{\eps_h}^2(|\vk+\vk'|)} \eps_h(\vk) \eps_h(\vk') \eps_h(-\vk-\vk')
  \Bigg]\,,
\nonumber
\ea
where $\sigma_{\eps_h}^2(k) =  P^\eps_{hh}(k)/\Lbox^3$. As an approximation to the proper marginalization over the stochastic fields, we now insert $\eps_h(\vk) = \d_h(\vk) - \d_{h,\rm det}(\vk)$ to obtain the leading correction to the likelihood $\ln P(\vec{\d}_h | \vec{\d}, \{b_O\}, \{\lambda_a\})$. Taking the expectation value of the derivative of this likelihood with respect to $b_O$, we obtain
\ba
\fsum{\vk} \frac1{\s^2(k)}\< [O](\vk) \d_h(\vk)\>  =\:& \sum_{O'} \hat b_{O'} \fsum{\vk} \frac1{\s^2(k)} \< [O](\vk) [O'](\vk)\> \qquad\forall\  O \vs
&\quad \stackrel{O=\d}{+} \left[ V^{\eps,2}_{hm} k^2 + V^{\eps,4}_{hm} k^4 + b_1 V^{\eps,4}_{mm} k^4 \right] \vs
& + \fsum{\vk} \fsum{\vk'} \frac{B_{\eps_h}(\vk,\vk') B_{\eps_h\eps_h O}(\vk,\vk')}{\Lbox^{12} \sigma_{\eps_h}^2(k)\sigma_{\eps_h}^2(k')\sigma_{\eps_h}^2(|\vk+\vk'|)}
\,,
\label{eq:maxlikeNG}
\ea
where the first two lines come from the Gaussian part and are identical to \refeq{maxlikeGFourier2}, while the third line is the approximate expression for the leading non-Gaussian correction. The latter involves the continuum-normalized cross-bispectrum between $\eps_h$ and $O$:
\be
B_{\eps_h\eps_h O}(\vk,\vk') \equiv \< \eps_h(\vk) \eps_h(\vk') O(-\vk-\vk') \>'\,.
\label{eq:BeeO}
\ee
Note that the only operator for which this contribution can be valid on large scales is $O= \d$, since, for any quadratic or higher-order operator, \refeq{BeeO} starts at 1-loop order (i.e. involves at least 3, rather than 2 power spectra), and thus has to be suppressed on large scales. The bispectrum $B_{\eps_h\eps_h \d}(\vk,\vk')$ has a straightforward interpretation: it corresponds to the modulation of the halo noise amplitude by a density perturbation $\d(|\vk+\vk'|)$.

In order to gain quantitative insight, let us thus consider $O=\d$ and assume that $\eps_h$ follows a Poisson distribution. We then have (see e.g. \cite{schmidt:2016a} and Sec.~4.1 of \cite{biasreview})
\ba
P^\eps_{hh}(k) &= \frac1{\bar n} \quad\Rightarrow\quad \sigma_{\eps_h}^2(k) = \frac1{\bar n \Lbox^3} \vs
B_{\eps_h}(\vk, \vk') &= \frac1{\bar n^2} \vs
B_{\eps_h\eps_h \d}(\vk, \vk') &= \frac{b_1}{\bar n} \Plin(|\vk+\vk'|) \,,
\ea
where $\bar n$ is the mean comoving number density of halos. Crucially, while the exact amplitude is not expected to match the actual noise field of halos, the leading scaling with $k$ of all these contributions holds regardless of the Poisson assumption. The final ingredient needed for the estimate is the number of Fourier modes included in the likelihood. This is, approximately,
\be
\frac{4\pi}{3} \frac{k_{\rm max}^3}{k_F^3} = \frac{4\pi}{3 (2\pi)^3} (\Lbox k_{\rm max})^3\,.
\ee
We then obtain the following upper limit on the size of the non-Gaussian contribution to the maximum-likelihood equality \refeq{maxlikeNG} at fixed $\vk$:
\ba
\left|\fsum{\vk'} \frac{B_{\eps_h}(\vk,\vk') B_{\eps_h\eps_h O}(\vk,\vk')}{\Lbox^{12} \sigma_{\eps_h}^2(k)\sigma_{\eps_h}^2(k')\sigma_{\eps_h}^2(|\vk+\vk'|)}\right|
&= b_1 \fsum{\vk'} \frac{(\bar n \Lbox^3)^{3}}{\Lbox^{12} \bar n^3 } \Plin(|\vk+\vk'|) \vs
&\lesssim
\frac{b_1}{6 \pi^2} k_{\rm max}^3 \Plin(k_{\rm max})\,.
\label{eq:NGestimate}
\ea
Again, the prefactor can differ by order unity, as it relies on the Poisson assumption. Crucially, the scaling with $k_{\rm max}$ is robust. \refeq{NGestimate} states that the non-Gaussianity of the halo noise field can be safely neglected as long as $k_{\rm max}^3 \Plin(k_{\rm max}) < \sigma_\Lambda^2$ is much less than one. We see that this is the same condition which is required for a reliable EFT likelihood in general.

Finally, let us turn to the non-Gaussianity of the matter noise field $\eps_m(\vk)$. As in the case of its variance, its non-Gaussian correlators are suppressed by powers of $k$. We roughly expect (see \refsec{PT:quadrnoise}) that $B_{\eps_m}(\vk,\vk') \sim k^2 k'^2 |\vk+\vk'|^2/ \knl^{12}$. The dominant contribution on large scales is expected to be the cross-correlation with the halo noise, whose magnitude can be roughly bounded to
\be
| B_{\eps_m \eps_h\eps_h}(\vk,\vk')| \lesssim \frac{k^2}{\knl^2} B_{\eps_h}(\vk,\vk')\,.
\ee
We thus expect that the contributions involving $\eps_m$ to the maximum-likelihood point \refeq{maxlikeNG} (whose precise form we have not derived here), are correspondingly suppressed by powers of $(k/\knl)^2$, and hence smaller than the halo noise contributions (unless one were to consider an extremely dense halo sample, or matter itself as tracer). This is the same conclusion as we have reached in the Gaussian case, where the leading contribution to the variance is $P_{hh}^\eps$, while $P_{hm}^{\eps}$ is only relevant at NLO.

\section{Relation to BAO reconstruction}
\label{sec:reco}

The linear matter power spectrum contains an oscillatory feature, the baryon acoustic oscillation (BAO) feature, induced by sound waves in the baryon-photon fluid before recombination \cite{eisenstein/hu}. Since the physical scale corresponding to this feature, the sound horizon at recombination $r_s$, is known, a measurement of its apparent scale in the clustering of galaxies allows for direct estimates of the angular diameter distance and Hubble rate as functions of redshift.

The BAO feature in the power spectrum of the evolved density field is broadened due to the nonlinear growth of structure. This broadening degrades the precision with which the scale of the BAO feature can be determined in galaxy clustering. The dominant source of this broadening are displacements induced by large-scale modes \cite{eisenstein/seo/white:2007}. Fortunately, since galaxy displacements are unbiased at lowest order in derivatives (on large scales), a fact which is ensured by the equivalence principle, the broadening obtained in a given realization of the density field can be predicted robustly.

For this reason, BAO reconstruction approaches have been developed. Generally, these work by first estimating the displacement field using the galaxy density smoothed on a large scale (via \refeq{ZA}, in case of the Zel'dovich approximation), and then moving galaxies back to their initial positions using this estimated displacement. Since the first implementations of the method \cite{eisenstein/etal:2007,eisenstein/seo/etal:2007}, many refined versions have been presented  \cite{noh/etal:2009,tassev/zaldarriaga:2012,burden/etal:2015,schmittfull/etal:2015,wang/etal:2017,schmittfull/baldauf/zaldarriaga:2017,hada/eisenstein:2018}. What all current reconstruction methods have in common is the backward-modeling approach and the presence of a smoothing scale.

Now let us consider BAO reconstruction from the perspective of a perturbative Bayesian forward model.
The Eulerian position $\vx$ at which a given galaxy is observed can be related to the corresponding Lagrangian position, i.e. position in the initial conditions, through the displacement $\v{s}$:
\be
\vx = \xfl(\vq,\tau) = \vq + \v{s}(\vq,\tau)\,.
\ee
Since standard reconstruction methods are based on inferring large-scale displacements by assuming a linear bias relation, let us do the same here; we return to this issue below. Assuming a deterministic forward model, as we do throughout, the final posterior \refeq{post} then becomes
\ba
P\left(\vec{\d}_\text{in},\theta,b_1,\{\lambda_a\}\Big| \vec{\d}_h\right) =\:&
\mathcal{N}_P P_\text{prior}(\vec{\d}_\text{in}|\theta)
P\left(\vec{\d}_h - b_1 \vec{\d} \Big| \{\lambda_a\}\right)\,.
\label{eq:postlin}
\ea
Due to mass conservation of matter, the density field is directly related to the Lagrangian displacement. In Fourier space, this relation becomes (e.g., \cite{tassev:2013})
\be
\d(\vk) = \int_{\vq} \exp\left[-i \vk\cdot \left(\vq+\v{s}(\vq)\right)\right]\,,\quad k > 0.
\ee
Note that this relation does not ensure that $\d(\vk=0)=0$.
However, this is not relevant here as we use a Fourier-space likelihood that does not include $\vk=0$. Using the latter, \refeq{postlin} becomes
\ba
- \ln P\left(\vec{\d}_\text{in},\theta,b_1,\{\lambda_a\}\Big| \vec{\d}_h\right) =\:& \fsum{\vk} \Bigg[ \frac{|\d_\text{in}(\vk)|^2}{2\Plin(k|\theta)} \vs
  & + \frac{1}{2\sigma^2(k)} \left|\d_h(\vk) - b_1 \int_{\vq} \exp\left[-i \vk\cdot \left(\vq+\v{s}[\vec{\d}_\text{in}](\vq)\right)\right]\right|^2 \Bigg]\,.
\nonumber
\ea
Here, we have assumed for simplicity that the range in wavenumber space covered by the prior on $\d_\text{in}$ is the same as that of the likelihood involving $\d_h$, although this does not have to be the case in practice (and is not in our actual implementation; \refsec{results}).

In order to gain further analytic insight into how the posterior derived here works, let us assume the Zel'dovich approximation (ZA). We stress that in practice, our forward model never consists of this Zel'dovich approximation, but always more sophisticated forward models; for example, 2LPT or full N-body simulations. However, the ZA allows for a particularly simple illustration for how BAO reconstruction works when using a field-level inference approach. In the ZA, the Lagrangian displacement is evaluated at linear order:
\be
\v{s}(\vp) = \v{s}_\text{ZA}(\vp) \equiv -\frac{i\vp}{p^2} \d_\text{in}(\vp)\,.
\label{eq:ZA}
\ee
We then obtain
\ba
- \ln P\left(\vec{\d}_\text{in},\theta,b_1,\{\lambda_a\}\Big| \vec{\d}_h\right) &\stackrel{\text{ZA}}{=} \fsum{\vk} \Bigg[ \frac{|\d_\text{in}(\vk)|^2}{2\Plin(k|\theta)} \label{eq:lnPZA}\\
  & + \frac{1}{2\sigma^2(k)} \left|\d_h(\vk)
- b_1 \int_{\vq} \exp\left[-i \vk\cdot \left(\vq + \int_{\vp} \frac{-i\vp}{p^2} \d_\text{in}(\vp) e^{i\vp\cdot\vq} \right)\right]\right|^2 \Bigg]\,.
\nonumber
\ea
In order to obtain cosmological constraints, we now need to marginalize over the initial phases $\d_\text{in}(\vk)$ as well as $b_1$. Clearly, despite the very simple forward model, \refeq{lnPZA} is a highly complex non-Gaussian and non-separable posterior, and the result of the marginalization is far from obvious.

Nevertheless, if one expands the exponential on the right-hand side up to linear order in $\d_\text{in}$, the non-prior part of the posterior reduces to a sum over $[\d_h(\vk) - b_1 \d_\text{in}(\vk)]^2/2\sigma^2(k)$, i.e. precisely the expected Gaussian, linear-bias likelihood in the large-scale limit. We can thus assume that, after marginalization over $b_1$, the posterior will peak around the correct large-scale modes of the density field.\footnote{If the set of cosmological parameters includes the power spectrum normalization $\sigma_8$, then there is a perfect degeneracy of $b_1$ and $\sigma_8$ in the linear regime. However, as discussed in the previous sections, this degeneracy is broken when the forward model and likelihood are consistently extended beyond linear order. Current reconstruction approaches instead adopt a prior on $\sigma_8$.} Let us thus assume that marginalizing over the initial phases with $|\vk| < k_c$, where $k_c$ will be determined below, fixes the amplitudes $\d_\text{in}(\vk)$ to their true values.

The BAO feature is an oscillatory feature in the linear power spectrum,
\be
\Plin(k|\theta=\{\theta', r_s\} ) = \Plin^\text{smooth}(k|\theta') \left[1 + A_\text{BAO} \sin(k r_s)\right]\,,
\ee
where $r_s$ is the sound horizon, and $\Plin^\text{smooth}(k)$ is the smooth (non-wiggle) part of the power spectrum, while $A_\text{BAO}$ is the amplitude of the BAO feature which is not important here. We are interested in the constraint on $r_s$ obtained from the posterior \refeq{lnPZA}. Again, we cannot explicitly marginalize over $\d_\text{in}$. We however see that, upon taking a derivative of the posterior with respect to $r_s$, in order to obtain the maximum posterior and the curvature around it, we obtain a sum over wavenumbers weighted by an oscillatory function of $k$ with frequency $r_s$.

The modes which dominate the BAO broadening effect are on much larger scales than the BAO feature. Hence, we choose $k_c \ll 1/r_s$, and marginalize over the modes with $|\vk| < k_c$. Following the discussion above, we then obtain
\ba 
& - r_s \frac{\partial}{\partial r_s} \ln P\left(\vec{\d}_\text{in},\theta,b_1,\{\lambda_a\}\Big| \vec{\d}_h\right) \stackrel{\text{ZA,marg.}}{=} \sum_{k_c<|\vk|<k_{\rm max}} (k r_s) \cos(k r_s) \label{eq:lnPZABAO}\\
&\qquad\qquad\times \left[ \frac{|\d_\text{in}(\vk)|^2}{2\Plin(k|\theta)}
+ \frac{1}{2\sigma^2(k)} \left|\d_h(\vk)
- b_1 \int_{\vq} \exp\left[-i \vk\cdot \left(\vq + \int_{\vp}^{k_c} \frac{-i\vp}{p^2} \d_\text{in}(\vp) e^{i\vp\cdot\vq} \right)\right]\right|^2 \right]\,.
\nonumber
\ea
Note that, due to the oscillatory nature of the BAO feature, the displacement terms on the right-hand side are enhanced by $k/p \sim 1/(r_s p)$, and thus significant.
\refeq{lnPZABAO} precisely corresponds to constraining the BAO scale by evaluating a Gaussian, linear-bias likelihood for the linear density field displaced to the Eulerian position predicted by the large-scale displacements $(-i\vp/p^2) \d_\text{in}(\vp)$ which are reconstructed from the large-scale halo field $\d_h(\vp)$. Thus, our posterior, combined with a Lagrangian forward model, naturally includes BAO reconstruction. The key difference from standard reconstruction approaches is that \refeq{lnPZABAO}, in keeping with the entire Bayesian inference approach, employs a \emph{forward model} of the BAO displacements, where the data are compared to the displaced initial density field. The former approaches on the other hand attempt to \emph{move the data back} to the initial positions. We stress that \refeq{lnPZABAO} is a rough approximation intended to isolate the relevant aspects of the inference approach; in practice, the actual posterior for $r_s$ is obtained after proper marginalization over all initial phases.

Moreover, unlike the case in reconstruction approaches applied to the data, which explicitly introduce a smoothing scale $\sim 1/k_c$ used to reconstruct the displacement field from the observed galaxy density field, the posterior presented here self-consistently uses information from all scales $ < k_{\rm max}$ to infer the displacements (again, assuming the proper marginalization over all initial phases is performed).
In our numerical implementation, described in detail in \cite{paperII}, we go beyond the Zel'dovich approximation and use second-order Lagrangian perturbation theory (2LPT). This improves the accuracy of the displacement field $\v{s}[\vec{\d}_\text{in}]$. Further, the posterior correctly includes the subleading effects of large-scale modes on the BAO feature, which correspond to overdense regions effectively behaving like a curved universe with correspondingly different expansion history \cite{sherwin/zaldarriaga:2012}. The Zel'dovich approximation does not correctly predict the amplitude of this effect.
We emphasize again that the approach presented here is not rescricted to the 2LPT forward model, and full N-body simulations could be used instead. 
Finally, since the same forward model is used for all operators, the posterior presented here also includes BAO reconstruction at the bispectrum level.

\section{Results}
\label{sec:results}

Using numerical tests based on simulated data, we now demonstrate
the applicability of the EFT likelihood delevoped in this paper to practical
applications. Since an in-depth discussion of the details of
implementation and performance is beyond the scope of the paper, we
restrict this section to include a limited number of basic examples
and defer a more comprehensive analysis to an upcoming publication
\citep{paperII}.

In what follows, we present results obtained with the set of
cosmological N-body simulations described in
\cite{Biagetti:2016ywx}. They were generated with the smoothed
particle hydrodynamics code GADGET-2 \citep{Springel:2005} using
gravity only, with box size $L = 2000 \, h^{-1} \mathrm{Mpc}$ and
$1536^3$ matter particles of mass $M_\mathrm{part} = 1.8 \times
10^{11} \, h^{-1}M_{\odot}$, assuming a flat $\Lambda$CDM cosmology
and parameters $\Omega_\mathrm{m} = 0.3$, $n_\mathrm{s} = 0.967$, $h =
0.7$, and $\sigma_8 = 0.85$. Identifying spherical overdensities above
200 times the background density, the Amiga halo finder algorithm
\citep{Gill:2004} was then used to construct halo catalogs at redshift
$z = 0$. We will use these halos as mock tracers (in real space) to
verify if our method allows to recover unbiased estimates of the input
parameters.

To comply with the requirements summarized in \refsec{main_summary},
we first project the halo catalog, linearly evolved matter density,
and nonlinear matter density fields onto high-resolution grids
with $512^3$ cells using cloud-in-cell assignment.
Here, the nonlinear matter density field is constructed from a forward evolution using second-order Lagrangian perturbation theory (2LPT), which can also be used for efficient sampling of initial phases. 
  Using the density field constructed from the N-body particles results in similar results, since our analysis is using only fairly large scales. 
We subsequently
apply a low-pass filter $W_\L(\vk)$ to the data in Fourier space,
removing modes for which $k > \Lambda=0.1 \, h \ \mathrm{Mpc}^{-1}$. Starting
from the filtered density fields, we are then able to compute
(renormalized) representations of the operators $O$ as well as
evaluate their Fourier space correlators $\< O O'^* \>$
numerically. With a particular focus on the estimation of cosmological
parameters, we can proceed to derive maximum likelihood estimates for
the scaled bias parameters $\beta_O$ and $\s_{8}$ given by
\refeq{maxlikeGFourier2}, where we truncated the scale-dependent
variance at order $k^2$. While this set of equations is nonlinear in
the parameters (in particular in $\alpha \equiv \sigma_8/\sigma_{8,\rm fid}$), explicit solutions can be easily found by means of
standard computer algebra systems.

\twofig{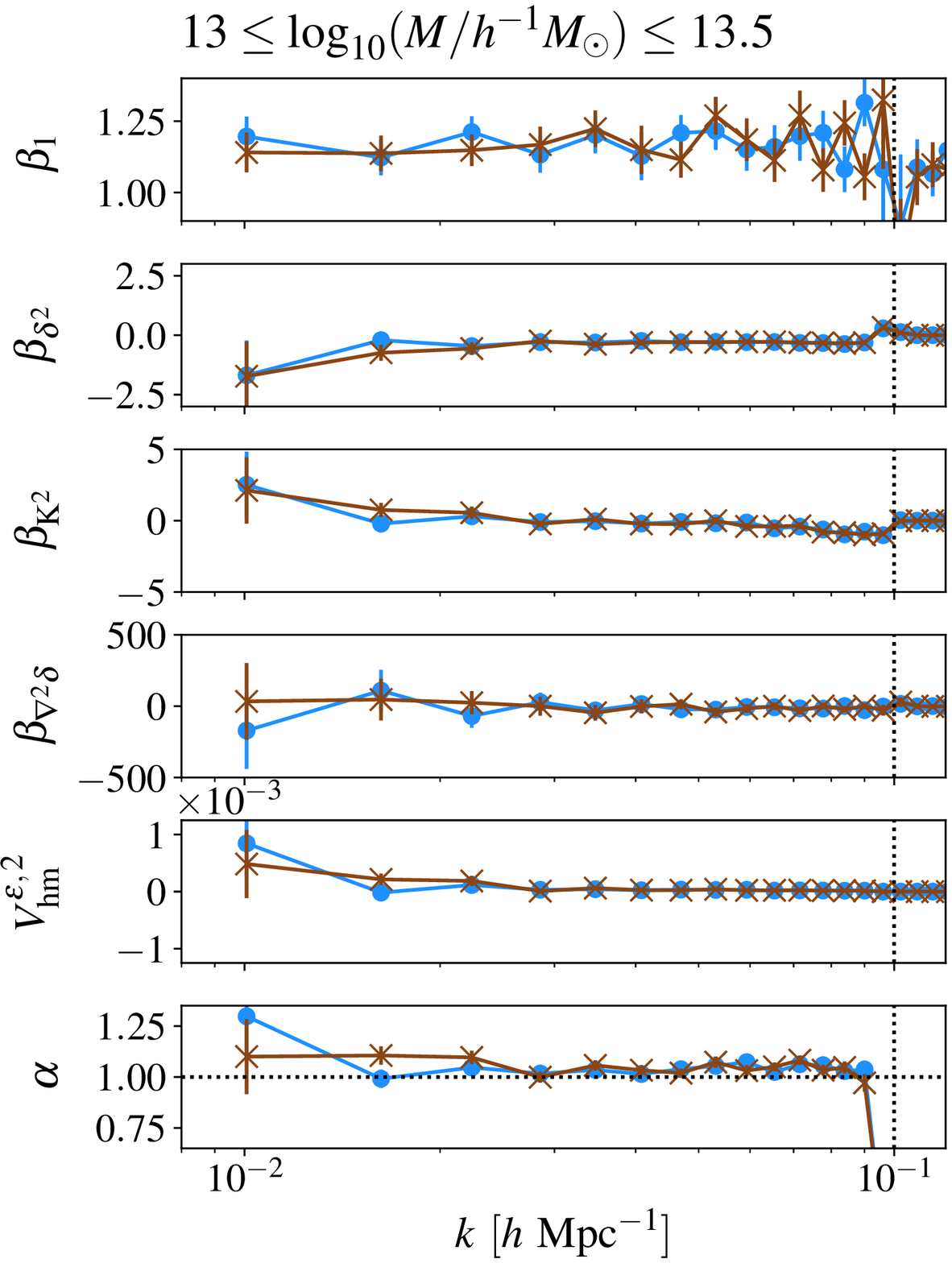}{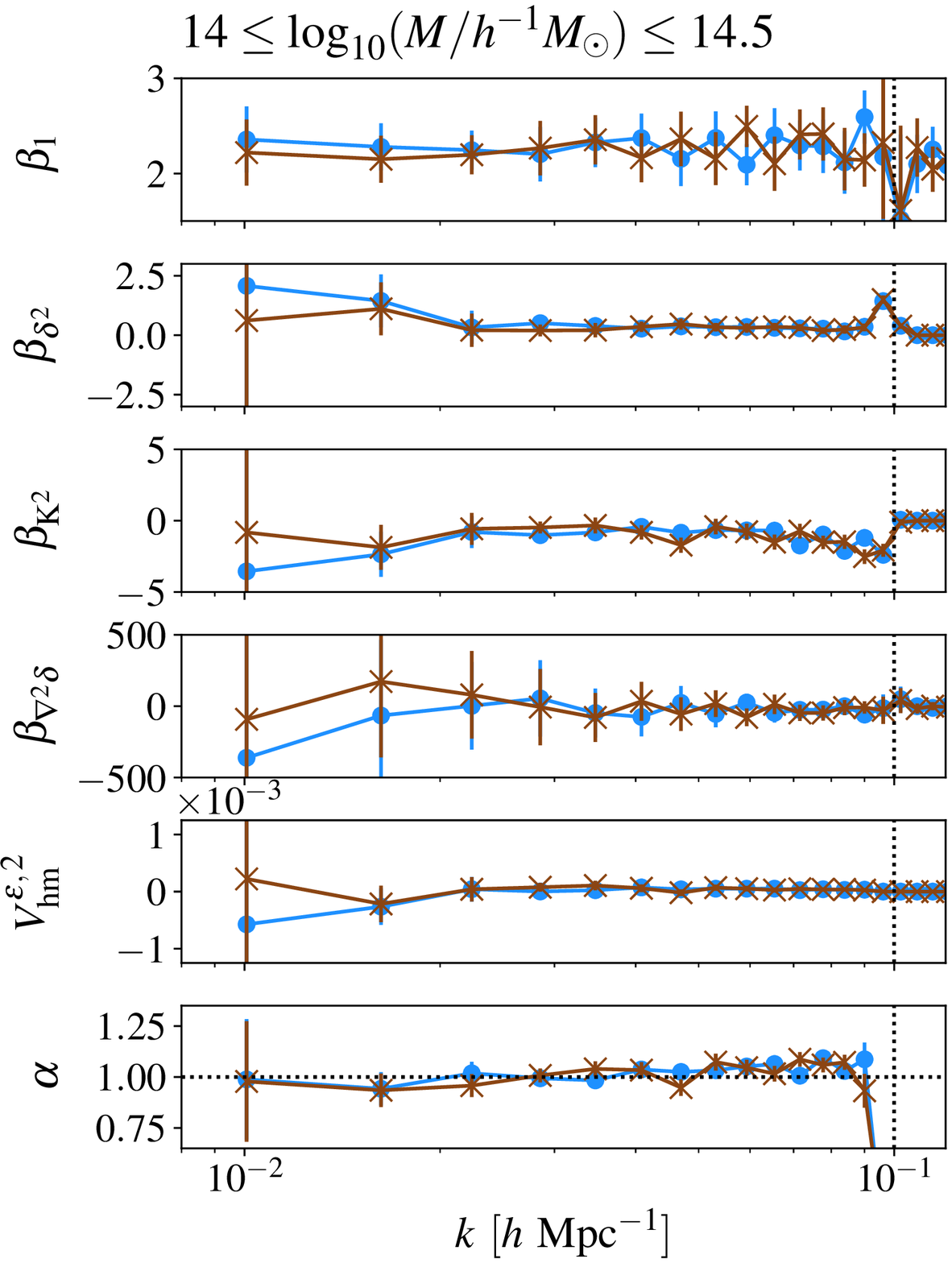}{fig:mle2step}%
{The EFT-based likelihood presented here allows for an unbiased measurement of
  bias and cosmological parameters like $\s_8$. For halos in the mass
  range $10^{13} \, h^{-1} M_\odot \le M \le 10^{13.5} \, h^{-1}
  M_\odot$ (\emph{left panel}) and $10^{14} \, h^{-1} M_\odot \le M
  \le 10^{14.5} \, h^{-1} M_\odot$ (\emph{right panel}), we show
  maximum likelihood estimates of the scaled bias parameters $\beta_1,
  \beta_{\delta^2}, \beta_{K^2}, \beta_{\nabla^2 \delta}$, the
  variance $V^{\epsilon, \, 2}_{hm}$, and $\alpha\equiv\sigma_8/\sigma_{8,\rm fid}$ (\emph{from top to
    bottom}) for two different simulations as a function of $k$. The
  filtering scale is indicated by vertical dotted lines while the
  horizontal dotted line in the bottom panels corresponds to an
  unbiased $\s_8$ estimate that matches the input value of the
  simulation. Error bars are 2-$\sigma$ bootstrap estimates over halo subsamples, which do not include residual cosmic variance.}

In \reffig{mle2step} we show the results of our analysis, plotted as a
function of wavevector $k$ in order to illustrate any systematic trends. Here, we are using
halos with masses $10^{13} \, h^{-1} M_\odot \le M \le 10^{13.5} \,
h^{-1} M_\odot$ (left panel) and $10^{14} \, h^{-1} M_\odot \le M \le 10^{14.5} \,
h^{-1} M_\odot$ (right panel) as biased tracers of the matter density field from two
independent simulation runs. Averaged over scales up to
$k_\mathrm{max} = 0.05 \, h \ \mathrm{Mpc}^{-1}$, we obtain numerical
values for the scaling parameters $\alpha^{13-13.5} = 1.03, \ 1.04$,
and $\alpha^{14-14.5} = 1.01, \ 0.99$ for the two realizations, respectively,
indicating that we can consistently recover the input parameters from
modes below the smoothing scale used in this example. The statistical
uncertainty in the inferred value of $\alpha$ is not simple to estimate,
since we have fixed the phases in our inference to the true values, and hence
cancel cosmic variance to the largest possible degree. The differences
between the two simulation realizations give a rough indication of the error
in the MLE parameters. Further, as a lower limit
on the errors associated with the estimation, we also show
uncertainty estimates from bootstrap samples of the halo catalog in \reffig{mle2step} but
stress that they fail to fully capture the cosmic variance
contribution. Clearly, $\sigma_8$ is correctly recovered up to a few percent accuracy, with some indication for a small positive bias.

We note in closing that the tests discussed here are in some sense the
most stringent possible probes of theoretical systematics, since a perfect knowledge of the underlying matter density
field has been assumed. In real-world applications, however, the phases have to
be inferred from the data and are subject to uncertainties that in
general result in larger parameter error bars, potentially making
remaining deficiencies in the theoretical modeling insignificant compared
to statistical uncertainties. We defer a detailed investigation of this question to
future work.

\section{Discussion and conclusions}
\label{sec:concl}

We have derived an EFT-based likelihood for the galaxy density field that
allows for cosmological inference from galaxy clustering with rigorously
controlled theoretical systematics, without the need for measuring arbitrarily
higher order $n$-point functions. In our concrete application, we have
restricted to a second-order bias expansion, including the leading higher-derivative bias contribution as well as scale-dependent stochasticity. While we have not included projection effects such as redshift-space distortions, and thus referred to the tracers as halos, the bias expansion is fully general and also holds for galaxies.

At this order, and when combined with a 2LPT forward model for matter, our posterior self-consistently combines the following sources of cosmological information in large-scale structure:
\begin{enumerate}
\item The leading- and next-to-leading order power spectrum, and leading-order bispectrum. In particular, this breaks the bias-amplitude degeneracy, which is perfect at linear order, using the second-order displacement term. Further, it allows for improved constraints on the slope (spectral index) of the linear power spectrum by extending the range in wavenumber $k$ useable for robust constraints.
\item Fully resummed BAO reconstruction using 2LPT displacements, both at the power spectrum and bispectrum levels.
\item Correct description of curvature and tidal effects on the local BAO scale. This effectively includes information from the 4-point function as well as higher-order statistics, through the highly nontrivial posterior in \refeq{post}.
\end{enumerate}
These probes translate into constraints on cosmological parameters. The first point allows for direct constraints on $\sigma_8$ and parameters which control the shape of the power spectrum, such as $\Omega_m$, $n_s$, and the sum of neutrino masses. The second and third allow for constraints on the expansion history and thus dark energy equation of state.
Quantifying the precise information content in terms of parameter constraints will be the subject of upcoming work.

We have also presented concrete numerical results validating the
theoretical derivation. Using halo catalogs obtained from N-body
simulations as physical tracers of the underlying matter density field, we
computed maximum-likelihood estimates of bias parameters and
$\sigma_8$. Evaluating the performance of our EFT likelihood approach at
different scales and halo mass ranges, we demonstrated that the input
parameter values can be consistently recovered. This is the first 
demonstration of unbiased cosmology inference for forward-modeling approaches
to date.

The likelihood can be straightforwardly extended to include higher-order bias contributions. While this might mean that non-Gaussian corrections to the likelihood and more noise fields need to be considered, the fundamental approach remains the same, and the dimensionality of the inference problem hardly changes. Thus, this approach appears much more feasible than explicitly measuring ever higher $n$-point functions.

Further, many additional types of cosmological physics can be included straightforwardly, such as the scale-dependent bias induced by primordial non-Gaussianity, as well as multiple tracers within the same volume. The main missing ingredient for the application to actual data are nontrivial survey geometries (mask) and redshift-space distortions. The former leads to a nondiagonal noise covariance  as discussed in \refsec{GaussF}, which however only needs to be determined once. 
Redshift-space distortions can be treated in the EFT approach as well \cite{perko/etal:2016,fonseca/etal:2018,pkgspaper}, and can be included in forward modeling correspondingly.\footnote{Redshift-space distortions allow for a measurement of the growth rate in the linear regime. However, if line-of-sight-dependent selection effects are present, then this information is removed due to a degeneracy with a new bias parameter. Then, nonlinear information is necessary to infer the growth rate \cite{pkgspaper}, in close analogy with the discussion in \refsec{sigma8} here.} We leave all of these developments to future work.

\acknowledgments

We would like to thank
Valentin Assassi,
Mikhail Ivanov,
Elisabeth Krause,
Marcel Schmittfull,
Marko Simonovi\'c,
and Matias Zaldarriaga, as well as the members of the Aquila Consortium,
for helpful discussions.
FE, MN, and FS acknowledge support from the Starting Grant (ERC-2015-STG 678652) ``GrInflaGal'' of the European Research Council.
GL acknowledges financial support from the ILP LABEX (under reference ANR-10-LABX-63) which is financed by French state funds managed by the ANR within the Investissements d'Avenir programme under reference ANR-11-IDEX-0004-02.  This work was supported by the ANR BIG4 project, grant ANR-16-CE23-0002 of the French Agence Nationale de la Recherche.
Part of this work was performed at Aspen Center for Physics, which is supported by National Science Foundation grant PHY-1607611.
This work is done within the Aquila Consortium.\footnote{\url{https://aquila-consortium.org}}

\clearpage
\appendix

\section{\texorpdfstring{Summary of EFT predictions for $\bm{n}$-point functions}{Summary of EFT predictions for n-point functions}}
\label{app:PTref}

Here we briefly summarize the relevant halo and matter $n$-point functions, which appear on the l.h.s. of the maximum-likelihood equality. These results are taken from Sec.~4.1 of \cite{biasreview}. For the halo-matter cross-power spectrum we have
\ba
P_{hm}^\NLO(k) =\:& b_1  P_{mm}^\NLO(k) + \hat{P}_{hm}^\NLO(k) \vs
\hat{P}_{hm}^\NLO(k) \equiv\:& 2 \sum_{O'\in \Oset^{[2]}} b_{O'} \int_{\vp} S_{O'}(\vp,\vk-\vp) F_2(\vp,\vk-\vp) \Plin(p) \Plin(|\vk-\vp|)
\vs
& + \left(b_{K^2} + \frac25 b_{\otd} \right) f_\NLO(k) \Plin(k) - b_{\lapl\d} k^2 \Plin(k) + k^2 P_{\eps \eps_m}^{\{2\}}\,,
\label{eq:PhmNLO}
\ea
where $f_\NLO(k)$ is defined as
\ba
f_\NLO(k) =\:& 4 \int_{\vp} \left[\frac{[\vp\cdot(\vk-\vp)]^2}{p^2 |\vk-\vp|^2}-1\right] F_2(\vk,-\vp) \Plin(p)\,,
\label{eq:fNLO}
\ea
and $F_2$ is given in \refeq{F2} below.
The matter and halo-matter-matter three-point functions at leading order are given by

\ba
B_{mmm}^\LO(k_1,k_2,k_3) =\:& 2 F_2(\vk_1,\vk_2) \Plin(k_1) \Plin(k_2) +
\perm{2}
\label{eq:Bmmm} \\
B_{hmm}^\LO(k_1,k_2,k_3) \equiv\:&  \< \d_h(\vk_1) \d(\vk_2) \d(\vk_3) \>'_{\LO}
\label{eq:Bhmm} \\
=\:& b_1 B_{mmm}^\LO(k_1,k_2,k_3)
+ \left[ b_2
+  2 b_{K^2} \left( \left[ \hat\vk_2\cdot\hat\vk_3 \right]^2 - \frac13\right) \right]
\Plin(k_2) \Plin(k_3)\,,
\nonumber
\ea
where the $F_2$ kernel is defined as
\ba
S_{\d^{(2)}}(\vk_1,\vk_2) \equiv F_2(\vk_1,\vk_2)
=\:& \frac57 + \frac27\frac{(\vk_1\cdot\vk_2)^2}{k_1^2k_2^2}
+ \frac{\vk_1\cdot\vk_2}{2k_1k_2}\left(\frac{k_1}{k_2}+\frac{k_2}{k_1}\right)
\label{eq:F2}
\ea

\section{Renormalized quadratic operators}
\label{app:renorm}

The renormalization conditions \refeq{renormcond} ensure that the cross-correlation of quadratic operators with the density field does in fact scale as a three-point function on large scales. Note that, $(\d_{W}^2)(\vk)$ involves a convolution in Fourier space. That is, small-scale (high-wavenumber) modes contribute to this term before renormalization, no matter how small $k$ is. Given the assumed Gaussian initial conditions, the leading contribution to $\< (\d_{W}^2) \d \>$ appears at second order in perturbation theory, and is obtained by replacing each instance of $\d_{W}$ with $\d_{W}^{(2)}$ (the third possible contribution vanishes). This yields (e.g., Sec.~2.10.4. of \cite{biasreview})
\ba
\< (\d_{W}^2)(\vk) \d(\vk') \>'\Big|_{\LO,k\to 0} =\:& 2 \< \left(\d_{W}^{(1)}\d_{W}^{(2)}\right)(\vk) \d^{(1)}(\vk')  \>' \vs
=\:& 2\int_{\vp_1}\int_{\vp_2}\int_{\vp_3}\!\!\! (2\pi)^3 \d_D(\vk-\vp_{123}) W(p_1) W(|\vp_{23}|) F_2(\vp_2,\vp_3) \vs
& \hspace*{2cm} \times \< \d^{(1)}(\vp_1) \d^{(1)}(\vp_2) \d^{(1)}(\vp_3) \d^{(1)}(\vk') \>' \vs
=\:& \Sigma_{1-3}^2(k) \, \Plin(k)\,,
\label{eq:d2dbare}
\ea
where
\ba
\Sigma_{1-3}^2(k) \equiv\:& 4\int_{\vp}
W(\vp) W(\vk-\vp)  F_2(-\vk,\vp) \Plin(p) \,.
\label{eq:Sigma13def}
\ea
We now specialize to a sharp-$k$ filter $W_\L(\vk)$.
We are interested in the large-scale limit $k\to 0$. If one were to approximate $W_\L(\vk-\vp) \approx W_\L(\vp)$, one would obtain $\Sigma_{1-3}^2(0) = \frac{68}{21} \< \d_\L^2\>$ (as usually done, e.g. \cite{assassi/etal,biasreview}). However, due to the sharp-$k$ filter, the actual low-$k$ limit is not quite that simple.

Let us look at \refeq{Sigma13def} in this limit in more detail. We have
\ba
\Sigma_{1-3}^2(k) =\:& \frac{4}{4\pi^2}\int_0^\Lambda p^2 dp  \Plin(p) \int_{-1}^1 d\mu\:
\Theta(\Lambda - |\vk-\vp|)  F_2(k,p, -\mu) \vs
=\:& \frac{1}{\pi^2}\int_0^{\Lambda-k} p^2 dp  \Plin(p) \int_{-1}^1 d\mu\: F_2(k,p, -\mu) \vs
& + \frac{1}{\pi^2}\int_{\Lambda-k}^\Lambda p^2 dp  \Plin(p) \int_{\mu_c(k,p,\Lambda)}^1 d\mu\: F_2(k,p, -\mu)\,,
\ea
where
\ba
\mu_c(k,p,\Lambda) = \frac{k^2+p^2-\Lambda^2}{2k p}\,.
\label{eq:muc}
\ea
If $k \ll \Lambda$, as relevant in the limit considered in \refeq{d2dbare}, these two contributions approximately become
\ba
\Sigma_{1-3}^2(k) \stackrel{k \ll \Lambda}{=}\:& \frac{68}{21} \sigma^2(\Lambda-k)
+ \frac{\Plin(\Lambda)}{4\pi^2} \frac{\Lambda}{k} \int_{\Lambda-k}^\Lambda p^2 dp
\left(1 - \mu_\text{min}^2\right) \vs
=\:& \frac{68}{21} \sigma^2(\Lambda-k)
+ \frac{\Lambda^3 \Plin(\Lambda)}{4\pi^2} \left[-\frac23 + \O \left( \frac{k}{\Lambda} \right) \right]
\,,
\ea
where we have kept only the leading term $\sim p/k$ inside $F_2$ in the low-$k$ limit. Finally, note that
\be
\s^2(\Lambda-k) - \s^2(\Lambda) = \O\left(\frac{k}{\Lambda}\right)\,.
\ee
This shows that there is an order unity correction to $\Sigma_{1-3}^2$ in the low-$k$ limit which remains even when $k/\Lambda \ll 1$, and we have
\be
\Sigma_{1-3}^2(0) = \frac{68}{21} \s^2(\Lambda) - \frac{\Lambda^3 \Plin(\Lambda)}{6\pi^2}\,.
\label{eq:Sigma130}
\ee
Generalizing the standard renormalization procedure, we could thus subtract the constant in \refeq{Sigma130}, multiplied by $\d(\vx)$, from $\d^2(\vx)$ to obtain the renormalized operator. However, in order to also remove the corrections that are linear in $k/\L$, we instead subtract the full function $\Sigma_{1-3}^2(k)$, as written in \refeq{renormfid}.

Finally, it is worth pointing out that the renormalization described here also removes a potentially worrisome contribution in the auto-correlation of quadratic operators. The auto-correlation of the \emph{un-renormalized} quadratic operators contains a connected contribution given by
\ba
\< O(\vk) O'(\vk') \>_c' &= \int_{\vp_1} S_O(\vp_1,\vk-\vp_1) \int_{\vp_2} S_O(\vp_2,\vk-\vp_2) T_{m,W}(\vp_1,\vk-\vp_1, \vp_2,-\vk-\vp_2)\,,
\ea
where $T_{m,W}$ denotes the matter trispectrum convolved multiplied by four filter functions, which at tree level involves terms of order $F_3$ and of order $(F_2)^2$. Let us consider the low-$k$ limit. The contributions involving $F_3$ depend only on the external momenta $\vp_i, \pm\vk-\vp_i$, which remain large (of order $\L$), and thus do not lead to a large correction in the low-$k$ limit. On the other hand, the terms of order $(F_2)^2$ involve sums of two momenta. In particular, there is a single type of diagram which has the form
\ba
\< O(\vk) O'(\vk') \>_{c,\LO}' &\supset 4\int_{\vp_1} S_O(\vp_1,\vk-\vp_1) \int_{\vp_2} S_O(\vp_2,\vk-\vp_2) 2 F_2(\vp_1,\vk) 2 F_2(-\vp_2,-\vk) \vs
& \hspace*{1cm}\times W_\L(\vp_1) W_\L(\vk-\vp_1) W_\L(\vp_2) W_\L(\vk-\vp_2) \Plin(p_1) \Plin(p_2) \Plin(k)\,,
\label{eq:trisplowk}
\ea
which scales roughly as $\sigma^4(\L) \Plin(k)$ and would thus amount to a significant correction at low $k$. The prefactor of 4 comes from the fact that we can shift each integration variable by $\vp_i \to \pm\vk - \vp_i$ to yield the same type of contribution. The factor 2 in front of each $F_2$ derives from the Feynman rules of standard perturbation theory.

However, this result applies to the un-renormalized operators. The corresponding correlator for the renormalized operators is given by
\ba
\< [O](\vk) [O'](\vk') \>_c' &= \< \left[O(\vk) - r_O \Sigma_{1-3}^2(k) \d(\vk)\right] \left[O'(\vk') - r_{O'} \Sigma_{1-3}^2(k') \d(\vk')\right]\>_c' \vs
&\stackrel{\LO,k\to 0}{\simeq} \< O(\vk) O'(\vk') \>_{\LO,c}' - r_O r_{O'} \left[\Sigma_{1-3}^2(k)\right]^2 \Plin(k)\,,
\ea
where we have used the result of the previous section for the correlator $\< O(\vk) \d(\vk')\>_{\LO,k\to 0}$. Comparing \refeq{trisplowk} with \refeq{Sigma13def}, we see that the leading contribution scaling as $\Plin(k)$ is canceled by renormalization. The remaining contributions to the connected correlator $\< [O](\vk) [O'](\vk')\>_c$ are suppressed at low-$k$ compared to the leading disconnected contribution derived in \refsec{PT:MLEO2}.

\section{\texorpdfstring{$\< \d_h \d \>$ and $\< O^{[2]} \d \>$}{< delta_h delta > and < O delta >}}
\label{app:opcorr}

In this appendix, we investigate the equality in \refeq{maxlikeGFourier2} for $O = \d_W$, \refeq{dlhsrhs} which we reproduce here, in more detail:
\ba
&
|W(\vk)|^2 \left[b_1 \Plin(k) + P_{hm}^\NLO(k) \right]\shouldeq
|W(\vk)|^2 \left\{ b_1 \left[\Plin(k) + P_{mm}^\NLO\right] - \clapl k^2 \Plin(k) \right\} \vs
& \hspace*{1.5cm}+ \sum_{O'\in \Oset^{[2]}} b_{O'} \bigg[ W(-\vk) \int_{\vp} S_{O'}(\vp,\vk-\vp) W(\vp) W(\vk-\vp) B_{mmm}(-\vk, \vp, \vk-\vp) \vs
 & \hspace*{3.8cm} -  r_{O'} \Sigma_{1-3}^2(k) |W(\vk)|^2 P_{mm}(k) \bigg] +  V^{\eps,2}_{hm} k^2\,.
\ea
For specificity, we assume in this appendix that the conditions derived for the quadratic operators in \refsec{PT:MLEO2} hold, i.e. that $W=W_\L$ is a sharp-$k$ filter with cutoff $\L$, and $W_\L(-\vk)=1$. Using \refeq{PhmNLO}, we can then simplify this relation to
\ba
&
\hat P_{hm}^\NLO(k) \shouldeq
\sum_{O'\in \Oset^{[2]}} b_{O'} \bigg[ \int_{\vp} S_{O'}(\vp,\vk-\vp) W_\L(\vp) W_\L(\vk-\vp) B_{mmm}(-\vk, \vp, \vk-\vp) \vs
  & \hspace*{3.8cm} -  r_{O'} \Sigma_{1-3}^2(k)  \Plin(k) \bigg]
- \clapl k^2 \Plin(k) +  V^{\eps,2}_{hm} k^2
\,.
\label{eq:dlhsrhsA}
\ea
On both sides of this equation, the terms involving momentum integrals can be separated into two components, which we refer to as ``2--2'' and ``1--3''. As we will see, the term $\clapl k^2 \Plin(k)$ can be associated with the 1--3 contributions, while $V^{\eps,2}_{hm} k^2$ is associated with the 2--2 terms.

\subsection{2--2 contributions}

The 2--2 contributions correspond to correlators of two quadratic operators. We have, on the l.h.s.,
\ba
\mbox{l.h.s.}_{2-2} =\:& 2 \sum_{O'\in \Oset^{[2]}} b_{O'} \int_{\vp} S_{O'}(\vp,\vk-\vp) F_2(\vp,\vk-\vp) \Plin(p) \Plin(|\vk-\vp|)\,,
\ea
and on the r.h.s.,
\ba
\mbox{r.h.s.}_{2-2} =
\sum_{O'\in \Oset^{[2]}} b_{O'} \int_{\vp} S_{O'}(\vp,\vk-\vp) W_\L(\vp) W_\L(\vk-\vp) 2 F_2(\vp,\vk-\vp) \Plin(p)\Plin(|\vk-\vp|) \,.
\ea
Note that we have included only one of the three permutations inside $B_{mmm}(-\vk,\vp,\vk-\vp)$ here.
We can then compute the difference between l.h.s. and r.h.s. as
\ba
\mbox{(l.h.s. -- r.h.s.)}_{2-2} =\:&
2 \sum_{O'\in \Oset^{[2]}} b_{O'} \int_{\vp} S_{O'}(\vp,\vk-\vp)
\left[1 - W_\L(\vp) W_\L(\vk-\vp)\right] F_2(\vp,\vk-\vp) \vs
& \hspace*{2.5cm}\times \Plin(p)\Plin(|\vk-\vp|)
\vs
=\:& 2 \sum_{O'\in \Oset^{[2]}} b_{O'} \int_{\L}^\infty \frac{p^2 dp}{(2\pi)^2} \int_{-1}^1 d\mu\: S_{O'}(\vp,\vk-\vp) F_2(\vp,\vk-\vp) \vs
& \hspace*{2.5cm}\times \Plin(p)\Plin(|\vk-\vp|) \vs
& +2 \sum_{O'\in \Oset^{[2]}} b_{O'} \int_{0}^\L \frac{p^2 dp}{(2\pi)^2} \int_{-1}^{\max\{-1,\mu_c(\L,k,p)\}} d\mu\: S_{O'}(\vp,\vk-\vp) F_2(\vp,\vk-\vp) \vs
& \hspace*{2.5cm}\times  \Plin(p)\Plin(|\vk-\vp|)
\,,
\label{eq:22residuals}
\ea
where $\mu_c$ is defined in \refeq{muc}. 
The first contribution in the last equality only involves momenta $p > \L$, where the integrand is dominated by modes $p \gg \L > k$. 
Since the integrand is analytic in $\vk$ around $k=0$, we can expand the integrand in a series of $k^2/p^2$. Moreover, the lowest-order, constant term vanishes, since $F_2(\vp,-\vp)=0$ (this follows from $\< \d^{(2)}(\vk) \>=0$). Thus, the leading contribution to the difference between l.h.s. and r.h.s. is of the form
\ba
\mbox{(l.h.s. -- r.h.s.)}_{2-2} =\:& A k^2\,,\quad\mbox{where}\quad |A| \sim \frac{1}{\L^2 \knl^3}\,.
\label{eq:Aksquare}
\ea
The magnitude of the constant $A$ was derived from power-law scaling arguments, but a numerical integration of the first term in \refeq{22residuals} confirms this (in fact, $A < 0$). 
The second line in \refeq{22residuals}, on the other hand, involves momenta $\vp$ such that $p < \L$ but $|\vk-\vp|> \L$. As long as $k < \L$, this also scales as $k^2$ to fairly high accuracy, as confirmed by numerical integration. 

Thus, the error of the 2--2 type can be absorbed by including a noise cross-power spectrum that scales as $k^2$, i.e. the term $V^{\eps,2}_{hm} k^2$ in \refeq{sigmak} and \refeq{dlhsrhs}.

\subsection{1--3 contributions}

The other type of contribution to the correlators involving the density field is the cross-correlation between linear and cubic operators. The latter are present in $\d_h$, although we have not included them in our bias expansion \refeq{dhdet}. We have, on the l.h.s., via \refeq{PhmNLO}
\ba
\frac{\mbox{l.h.s.}_{1-3}}{\Plin(k)} =
\left(b_{K^2} + \frac25 b_{\otd} \right) f_\NLO(k) - b_{\lapl\d} k^2 \,.
\label{eq:dRcorrlhs13}
\ea
On the r.h.s., we have the remaining two permutations inside $B_{mmm}$ not included under the 2--2 contributions, as well as the counterterms and the higher-derivative bias, which yield
\ba
\frac{\mbox{r.h.s.}_{1-3}}{\Plin(k)} =\:& 2 \sum_{O'\in \Oset^{[2]}} b_{O'} \bigg\{\int_{\vp} S_{O'}(\vp,\vk-\vp) W_\L(\vp) W_\L(\vk-\vp) \vs
&\hspace*{3.2cm}\times \left[ F_2(-\vk,\vp) \Plin(p)
  + F_2(-\vk,\vk-\vp) \Plin(|\vk-\vp|)\right]
\vs
& \hspace*{2cm} -  r_{O'} \Sigma_{1-3}^2(k) \bigg\} - \clapl k^2
\,.
\ea
First, notice that this corresponds to a contribution to the correlator $\< \d(\vk) O'^{[2]}(\vk')\>'$ which is proportional to $\Plin(k)$, so one can expect the counterterms in the third line to become relevant.  The integrand is symmetric under interchange of $\vk \leftrightarrow \vp-\vk$, and this applies to the factor outside of the square brackets by itself as well. By shifting the integration variable, one easily sees that the second term in brackets leads to the same result as the first.

Inserting the explicit expressions for the kernels $S_O$, using that $S_{K^2}(\vp_1,\vp_2)=\mu_{\vp_1,\vp_2}^2 - 1 + 2/3$, we obtain
\ba
\frac{\mbox{r.h.s.}_{1-3}}{\Plin(k)} =\:&  4 b_{K^2} \int_{\vp} \left[\frac{[\vp\cdot(\vk-\vp)]^2}{p^2 |\vk-\vp|^2}-1\right] W_\L(\vp) W_\L(\vk-\vp) F_2(-\vk,\vp) \Plin(p) - \clapl k^2 \vs
& + \left\{ b_{\d^2} + \frac23 b_{K^2}\right\} \left[ 4  \int_{\vp} W_\L(\vp) W_\L(\vk-\vp) F_2(-\vk,\vp) \Plin(p)  -  \Sigma_{1-3}^2(k) \right]
\,.
\label{eq:dRcorrrhs13}
\ea
We then obtain for the difference between l.h.s. and r.h.s. of \refeq{dlhsrhsA}
\ba
\frac{\mbox{r.h.s.--l.h.s.}_{1-3}}{\Plin(k)} =\:&
 4 b_{K^2} \int_{\vp}  \left[\frac{[\vp\cdot(\vk-\vp)]^2}{p^2 |\vk-\vp|^2}-1\right] W_\L(\vp) W_\L(\vk-\vp) F_2(-\vk,\vp) \Plin(p) \vs
  & - \left(b_{K^2} + \frac25 b_{\otd} \right) f_\NLO(k) + \left(b_{\lapl\d}-\clapl\right) k^2  \,,
\label{eq:dRcorrres13}
\ea
where we have used \refeq{Sigma13def}. Note that the two bispectrum permutations which scale as $\Plin(k)$ have been canceled completely by the counterterms in \refeq{renormfid}.

Let us first disregard the last term, $\propto k^2$. Comparison with \refeq{fNLO} shows that, if $b_\otd$ were zero, the residual would again be entirely due to the presence of the smoothing kernel in the loop integral, as in the case of the 2--2 contributions discussed previously. In this case, since the loop integral in \refeq{fNLO} is again dominated by modes $p\gg \L$, the residual is proportional to $k^2 \Plin(k)$, and can be absorbed by $\clapl$. This motivates our notation $\clapl$, since this parameter is not expected to be equal to the actual higher-derivative bias $b_{\lapl\d}$ due to the presence of other corrections.

However, a further residual is induced by the cubic bias parameter $b_{\otd}$, which the r.h.s. does not contain, since our bias expansion only includes terms up to quadratic order. However, as long as one can accurately approximate $f_\NLO(k)$ as $f_\NLO(k) \simeq c k^2$, this cubic bias contribution can be absorbed into the effective higher-derivative bias $\clapl$ as well. This approximation is very accurate as long as $k \ll \knl$. Specifically, for a $\Lambda$CDM cosmology and for $k \lesssim \Lambda/2$, the fractional deviation from a pure $k^2$ is on the order of 5\%.

To conclude, we find that the maximum-likelihood point for the Fourier-space likelihood [\refeq{maxlikeGFourier2}] holds for $O=\d_\L$, even when including all NLO corrections, provided the conditions listed in \refsec{GaussF} hold, and one allows both the higher-derivative bias $b_{\lapl\d}$ and the scale-dependent noise $V^{\eps,2}_{hm}$ to vary. The latter two parameters effectively absorb all NLO corrections that are not explicitly contained in the bias model.

\section{Maximum-likelihood estimate for variance}
\label{app:sigmak}

Let us write $\s^2(k)$ in \refeq{PcondGFT} as
\be
\s^2(k) = \sum_{n=0,2,4} \s^2_n k^n\,.
\ee
The maximum-likelihood points for the parameters $\sigma_n^2$ are then given by
\be
0 \shouldeq \fsum{\vk} k^n \left[\frac12 \hat\s^{-2}(k) - \hat\s^{-4}(k) \left\{ \d_h(\vk) - \d_{h,\rm det}(\vk)\right\}^2 \right]\,,\qquad
n = 0, 2, 4\,.
\ee
That is, $\s^2(k)$ is attempting to match the correlator
\be
\< \left|\d_h(\vk) - \d_{h,\rm det}(\vk) \right|^2 \> = 2 \sum_{n=0,2,4,\cdots} \hat\sigma^2_n k^n\,,
\label{eq:sigmaMLE}
\ee
where $\hat\sigma_n$ are the maximum-likelihood values for $\sigma_n$.
Note that the expansion on the right-hand side only holds if all relevant contributions to $\d_h(\vk)$ from modes $k < \Lambda$ are included in $\d_{h,\rm det}(\vk)$. It is easy to see that this holds for the likelihood derived in this paper: \refeq{sigmaMLE} corresponds to the mean squared residuals of the maximum-likelihood equations \refeq{maxlikeGFourier2}. There are two sources of such residuals: actual noise in the halo density field, described by the statistics of the noise field $\eps_h$, and residuals due to the imperfect forward modeling.
As we have shown in \refsec{PT:MLEO2} and \refapp{opcorr}, the latter are of the form $A k^2$ [cf. \refeq{Aksquare}] and $c k^2 \Plin(k)$. The second type of residual is proportional to the long-wavelength modes themselves, and is thus removed by including the term $\clapl k^2 \d(\vk)$ in $\d_{h,\rm det}(\vk)$. The first type corresponds to true stochastic residuals which contribute to the effective noise. We can thus write
\be
\hat\sigma^2_n = V_{hh}^{\eps,n} + V_\text{eff}^{\eps,n}\,,
\ee
where $V_\text{eff}^{\eps,0}=0$, while, from \refeq{Aksquare}, we expect $V_\text{eff}^{\eps,2} \sim (k^2/\L^2) \knl^{-3}$. Note that $V_{hh}^{\eps,0}$ corresponds to the true large-scale halo shot noise, while higher-order noise variances are to be considered effective parameters as they absorb stochastic errors in the model. Further, by including both $V_{hh}^{\eps,2}$ and $V_{hm}^{\eps,2}$ in the likelihood, we guarantee that the mean as well as the residuals of the MLE equations can be absorbed consistently. 
To the order we work in throughout this paper, the contributions of order $k^4$ and higher to the variance can be neglected.

Note that this reasoning does not strictly apply to the higher-loop and higher-derivative contributions discussed in \refsec{PT:HO}, since they do not scale analytically in $k$. Thus, for consistency, one could also include a template for higher-order corrections in the noise variance $\sigma^2(k)$.

\bibliographystyle{JHEP}
\bibliography{bibliography}

\end{document}